\documentclass[aps,prb,twocolumn,amssymb,superscriptaddress]{revtex4}
\usepackage{amssymb}
\usepackage{amsmath}
\usepackage{graphicx}
\usepackage{epsfig}

\usepackage{graphicx}

\def\be{\begin{equation}}
\def\ee{\end{equation}}
\def\ba#1{\begin{array}{#1}}
\def\ea{\end{array}}
\def\bn{\begin{enumerate}}
\def\en{\end{enumerate}}

\def\r{\right}
\def\l{\left}

\def\zs{\zeta_D}

\def\scs{SC$^*$}

\def\lb#1{\label{#1}}

\usepackage[latin1]{inputenc}

\begin{document}

\title{Simulation results for an interacting pair of resistively shunted Josephson junctions}
\author{Philipp Werner}
\affiliation{Institut f{\"u}r theoretische Physik, ETH H{\"o}nggerberg, CH-8093 
Z{\"u}rich, Switzerland}
\author{Gil Refael}
\affiliation{Kavli Institute of Theoretical Physics, University of California, Santa Barbara, CA 93106}
\affiliation{Department of Physics, California Institute of Technology, Pasadena, CA 91125}
\author{Matthias Troyer}
\affiliation{Institut f{\"u}r theoretische Physik, ETH H{\"o}nggerberg, CH-8093 
Z{\"u}rich, Switzerland}

\date{\today}

\begin{abstract}

Using a new cluster Monte Carlo algorithm, we study the phase diagram and critical properties of an interacting pair of
resistively shunted Josephson junctions. This system models tunneling  between two
electrodes through a small superconducting grain, and is described by a double sine-Gordon
model \cite{Refael1}. In accordance with theoretical predictions, we observe
three different phases and crossover effects arising from an
intermediate coupling fixed point. On the superconductor-to-metal
phase boundary, the observed critical behavior is within error-bars
the same as in a single junction, with identical values of the
critical resistance and a correlation function exponent which depends
only on the strength of the Josephson coupling. We explain these
critical properties on the basis of a renormalization group (RG) calculation. In addition, we propose an
alternative new mean-field theory for this transition, which correctly predicts the
location of the phase boundary at intermediate Josephson coupling
strength.

\end{abstract}

\maketitle
\thispagestyle{empty}

\section{Introduction}

The effects of dissipation and decoherence are ubiquitous in quantum systems 
and influence the properties of materials and
nano-scale devices in a profound way. Already the simplest model
system, an Ising spin in a transverse field which is coupled to an Ohmic
heat bath, displays interesting behavior such as a dynamical phase
transition to a localized state at a critical value of the dissipation
strength \cite{Chakravarty}. Another prominent example is the 
resistively shunted Josephson junction, which undergoes a
superconductor-to-metal transition at a critical value of the shunt
resistance \cite{Schmid, Bulgadaev, Guinea, Fisher&Zwerger}, which
equals the quantum of resistance $R_Q = h/4e^2=6.5\text{k}\Omega$. 
Arrays of Josephson junctions with dissipation have been studied both as a model for granular superconducting
films or nano-wires \cite{Chakravarty&Ingold1, Panyukov&Zaikin,
  Korshunov1}, and in their own right \cite{Haviland,
  Kobayashi}. The behavior of these systems, and in particular their
superconductor-to-metal 
phase transition, is far from being
completely understood. 

Recently, Refael \textit{et al.} have studied a model of a mesoscopic
superconducting grain connected to two leads via Josephson
tunneling and shunt resistors \cite{Refael1}. The phase diagram for this simple
system was shown to be remarkably complex, with three distinct phases.
In addition, contrary to the case of a single resistively shunted
Josephson junction, one of the phase boundaries is controlled in part by an intermediate coupling fixed
point, and the superconductor-to-normal transition across this boundary can be
tuned by the Josephson energy itself. The above effects indicate that
although the system discussed in Ref. \onlinecite{Refael1} is zero
dimensional, it nearly has the full complexity of a one-dimensional
{\it array} of Josephson junctions. Much of the recent results on
Josephson junction chains in Ref.~\onlinecite{Refael2} draw directly from
the two-junction system. The RG-flow equations of the two-junction system are also nearly identical to those of the two-dimensional triangular lattice presented in
Ref.~\onlinecite{Tewari}. In this closely related work on Josephson
junction arrays, the local nature of the phase
transition, as well as floating phases have been discussed. 

\begin{figure}[t]
\begin{minipage}[t]{\linewidth}
\centering\epsfig{figure=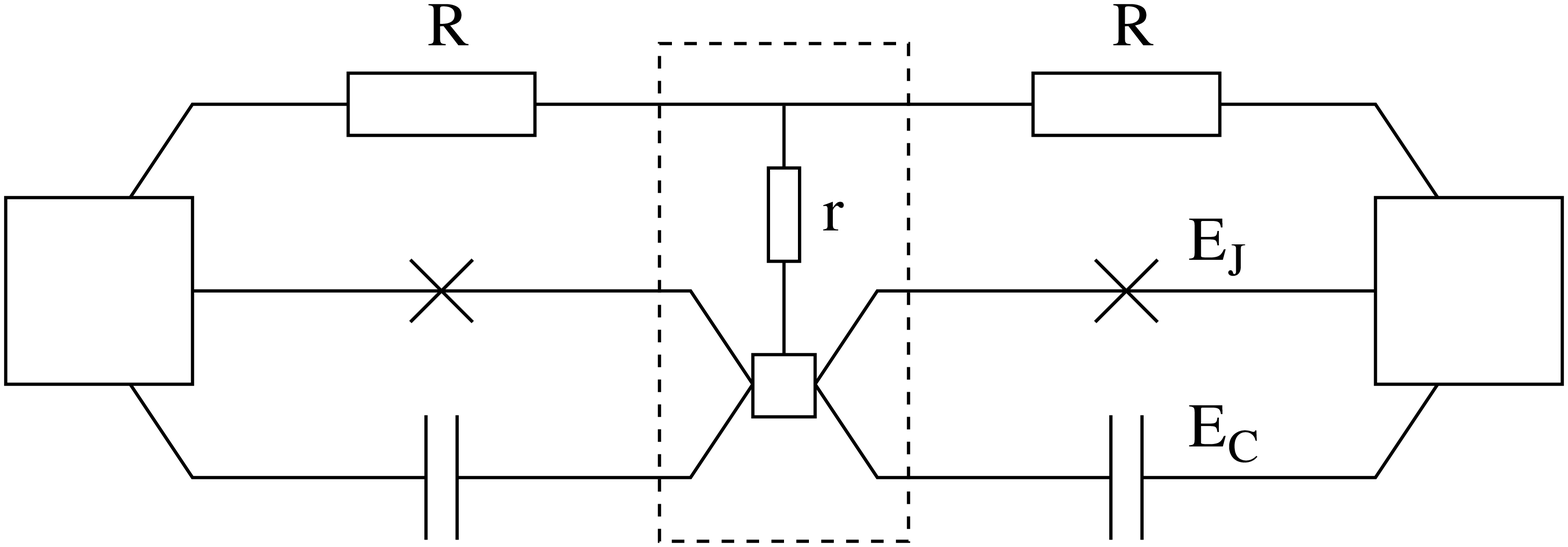,angle=0,width=8cm}
\end{minipage}\\
\vspace{3mm}
\begin{minipage}[b]{\linewidth}
\centering\epsfig{figure=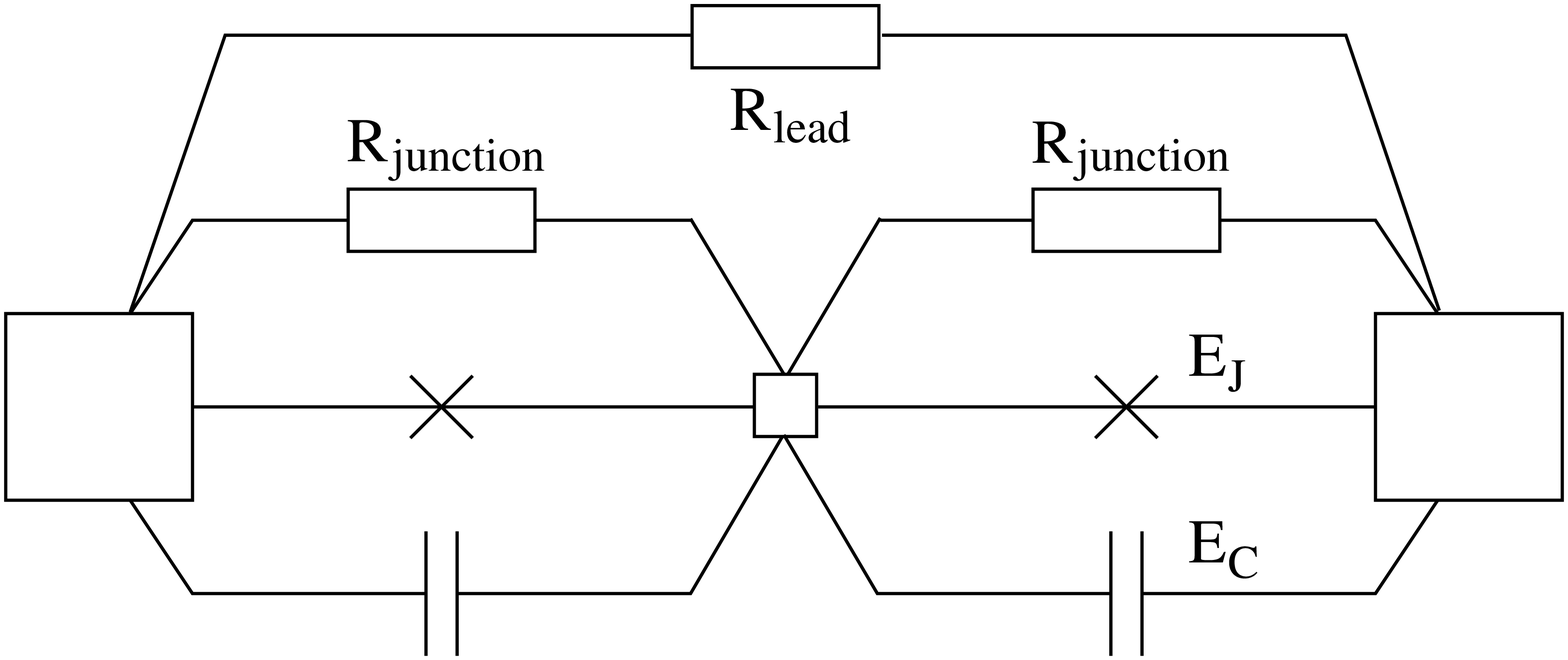,angle=0,width=8cm}
\end{minipage}
\caption{Upper figure: Two-junction model considered in
  Ref.~\onlinecite{Refael1} with identical shunt resistors $R$ for the
  left and right junction. The dotted box represents the central grain
  which incorporates a phenomenological charge relaxation mechanism
  described by the resistance $r$. \\
Lower figure: Equivalent model with modified shunt resistors $R_\text{junction}=R+2r$ 
and an additional resistor $R_\text{lead}=(R/r)(R+2r)$ connecting the leads.}
\label{two_junctions_refael}
\end{figure}

The development of a powerful new cluster Monte Carlo algorithm
\cite{junction} has allowed to test and verify several analytical
predictions for the single junction and to observe continuously
varying correlation exponents along the phase boundary. In this paper
we will use adapted versions of these cluster moves to simulate the
two-junction model of Ref.~\onlinecite{Refael1}, which is
the simplest extension of  the single junction case which exhibits
interesting new physics. The model (with identical shunt resistors) is shown in the upper part of
Fig.~\ref{two_junctions_refael}. It consists of two Josephson
junctions with coupling energy $E_J$, each shunted with an Ohmic
resistance $R$. On the central grain, the model incorporates a ``charge
relaxation mechanism'' (supposed to represent the break-up of Cooper
pairs into electrons) which is described by an additional
resistance $r$. 

Dissipation produced by electrons flowing through the resistors
reduces phase fluctuations between the superconducting islands which
they connect. This can be seen from the dissipative action term in
Eq.~(\ref{S_D}), and it explains how strong dissipation leads to
superconducting phase coherence. Depending on the values of $R$ and $r$, three different phases occur:
\begin{itemize}
\item The individual junctions are insulating and there is no
  super-current from lead to lead: Normal phase (NOR), 
\item The junctions are superconducting and thus also the whole device
  from lead to lead: Fully Superconducting phase (FSC),
\item The individual junctions are insulating but there is
  superconducting phase coherence from lead to lead: SC$^*$ phase.
\end{itemize}

\begin{figure}[t]
\centering
\includegraphics[width=8.5cm]{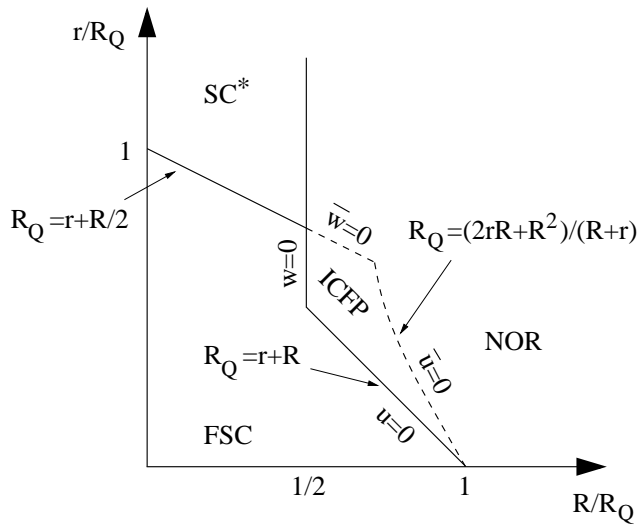}
\caption{Phase diagram in the limits $E_J/E_C \ll 1$ (solid lines) and
  $E_J/E_C \gg 1$ (solids and dashed lines). Besides the
  superconducting (FSC) and metallic (NOR) phase, the two-junction
  system can be in a state which is superconducting from lead to lead,
  although the individual junctions are insulating (\scs). The parameters $u,\,w$ and $\overline{u},\,\overline{w}$
  are defined in Eqs. (\ref{eq14.5}) and (\ref{i17}). The values $u=0$ and $w=0$
  define the FSC-NOR boundary in the weak coupling limit, and $\overline{u}=0$ and $\overline{w}=0$ in the strong coupling limit.
\label{phasediagram_refael}
\label{pdtheory}}
\end{figure}

The phase diagram for the limiting cases $E_J \ll E_C$ and $E_J \gg
E_C$ has been computed in Ref.~\onlinecite{Refael1} using an RG approach and is shown in
Fig.~\ref{phasediagram_refael} as a function of the
resistances $R$ and $r$. The boundary between the NOR- and FSC-phase
in the region marked ICFP depends on the value of $E_J/E_C$ as well as
the values of the resistors $r$ and $R$. The behavior of the system in this region
is controlled by an intermediate coupling fixed point
(ICFP), which will be discussed in Sec. \ref{ICFPth}. 

In the remains of this paper we present the results of a thorough
Monte Carlo (MC) investigation of the two-junction system. Each
result is compared with predictions or explanations based on the
RG flow equations presented in Sec. II. After discussing the various phases 
of the two-junction system, we explain in Sec. III the Monte Carlo method which was used
to investigate the model numerically. In particular, we discuss several types of efficient cluster updates, which are adaptations of the recently developed rejection-free cluster algorithm for single resistively shunted junctions.\cite{junction} 

In Sec. IV we identify the three phases NOR, FSC and SC$^*$
by computing the temperature dependence of the lead-to-lead and
lead-to-grain resistance.  The numerically obtained phase
diagram at intermediate Josephson coupling is compared to the theoretical
predictions for weak and strong Josephson coupling. There is a good agreement between theory
and simulation results, except in the region, where the three phases meet. We
explain these small deviations of the measured phase boundaries from
the predicted ones in terms of slow cross-overs in the RG flow, which
prevent the MC simulation from exploring the zero-temperature
behavior.

In Sec.~V we concentrate on the critical FSC-NOR line in the ICFP
region. In particular, we compare the critical properties of the
two-junction system to those of a single junction with the same Josephson
coupling $E_J/E_C$, and a shunt resistance $R_s=R_Q$. The two
systems exhibit (within error-bars) identical behavior in their
correlation functions, fluctuations and the effective resistance of
the critical junctions. These critical properties depend on the value
of $E_J/E_C$ but not on the value of the resistors $r$ and $R$.  In
Sec.~V~B, we try to explain this rather surprising observation in terms of the
effective junction resistances of the two-junction system, calculated from the RG flow of
Sec.~II. We show that the predicted critical resistance exhibits only
a weak dependence on the (shunt) resistors and agrees quite well with
the measured value.

In Sec. VI, however, we pursue a different path to explain the
remarkable resemblance of the single resistively shunted junction and
the two-junction system. We show that the data can be well explained
by a `mean-field' theory, which treats one of the two junctions as an
effective resistor. This is a new way of approaching the double
sine-Gordon model action (Eq. \ref{action1} below). The assumption is
that on the FSC-NOR phase boundary, each junction sees an environment
which imitates a shunt resistor $R_s=R_Q$ and that it may be replaced
by a resistor whose value can be found from the critical resistance of
the {\it single} resistively shunted junction (with the same Josephson
coupling). On the basis of these assumptions it is possible to derive
an expression for the position of the FSC-NOR phase boundary, which
fits the MC data and the RG-based predictions quite well.

In Sec.VI we consider the phase-phase correlation functions in the NOR
phase within the ICFP region. For fixed resistors, we measure a strong
dependence of the correlation exponents on the Josephson coupling
strength, which is in contrast to the single junction model, where
these exponents depend only on the value of the shunt resistor. This
behavior is explained as a consequence of the flow in the additional
Josephson coupling $J_+$ between the leads, which is generated under
the RG and not present in the single junction model.


\section{Theory for the symmetric two-junction system \label{theory}}

In this section we will present the effective action for the
symmetric two-junction system, and then briefly discuss its RG flow
equations and the various
phases. This discussion will prove to be especially useful when interpreting the
Monte Carlo results. 

We will first describe the NOR-\scs and \scs-FSC transitions in
the weak and strong coupling limits. 
It is important to note that the
\scs phase 
appears due to interactions between the two
junctions, and can not be understood in terms of the physics of a
single junction.
Finally, we will discuss the
important {\it intermediate coupling fixed point} which controls the
direct NOR-FSC transition. 

\subsection{Effective action}

The imaginary-time effective action of the symmetric two-junction system
can be written as a functional of the phase differences $\phi_1$ and
$\phi_2$ across the first and second junction,
\begin{eqnarray}
S_\text{eff}[\phi_1, \phi_2]  &=&  S_C[\phi_1, \phi_2] + S_J[\phi_1, \phi_2] + S_D[\phi_1, \phi_2],\hspace{2mm}
\lb{action1}
\end{eqnarray}
where the charging term $S_C$, the Josephson coupling term $S_J$ and the dissipation term $S_D$ read
\begin{eqnarray}
S_\text{C}[\phi_1, \phi_2]  &=& \frac{1}{16E_C}\int_0^\beta d\tau \Big[\Big( \frac{d\phi_1}{d\tau}\Big)^2 + \Big( \frac{d\phi_2}{d\tau}\Big)^2 \Big],\label{S_C}\\
S_\text{J}[\phi_1, \phi_2]  &=& -E_J\int_0^\beta d\tau[\cos(\phi_1)+\cos(\phi_2)],\label{S_J}\\
S_D[\phi_1,\phi_2] &=& \frac{R_Q}{R(R+2r)}\int_0^\beta d\tau d\tau' \frac{(\pi/\beta)^2}{\sin^2((\pi/\beta)(\tau-\tau'))}\nonumber\\
&\times&\big[R(\phi_1(\tau)-\phi_1(\tau'))^2 + R(\phi_2(\tau)-\phi_2(\tau'))^2\nonumber\\
&&  + r((\phi_1(\tau)+\phi_2(\tau))-(\phi_1(\tau')+\phi_2(\tau')))^2 \big].\label{S_D}\nonumber\\
\end{eqnarray}
$E_C=e^2/2C$ is the (single electron) charging energy of each junction
and sets the overall energy scale. $E_J$ denotes the coupling
strengths of the junctions. Ohmic dissipation in the resistors is
introduced using the model of Caldeira and Leggett \cite{Caldeira,
  Schoen}.
 
The system discussed in Ref.~\onlinecite{Refael1} is equivalent to
the one illustrated in the lower part of
Fig.~\ref{two_junctions_refael}, where each junction is shunted by a
resistor 
\be
R_\text{junction} = R+2r
\label{R_junction}
\ee
and the leads are connected by an
additional resistor 
\be
R_\text{lead}=(R/r)(R+2r). 
\label{R_lead}
\ee
This is a consequence of the ``Y-$\Delta$" transformation of resistor
networks \cite{Refael1}. For simplicity, we
consider a capacitive coupling between the leads and central grain
only and no Josephson coupling between the leads. Such a coupling will
be generated in the renormalization flow.

\subsection{Weak coupling limit}

\begin{figure}[t]
\includegraphics[width=8cm]{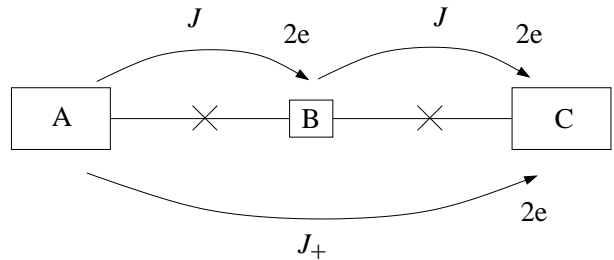}
\caption{In the weak-coupling limit, we consider the RG flow of the
  Josephson couplings $J$ (with $J_0=E_J$) and $J_+$. $J$ is the
  amplitude for Cooper-pair hopping between either of the leads and the grain. $J_+$ is the amplitude for
Cooper-pair tunneling from lead to lead. This process is generated in
the second order of the RG flow.
\label{weakfig1}}
\end{figure}

When the Josephson coupling energy $E_J$ is small, it can be used as a
small parameter for a perturbative
RG analysis. As explained in
Ref. \onlinecite{Refael1}, in addition to the bare Josephson energy of the
two junctions, the RG flow produces yet another Josephson coupling - the
cotunneling $J_+$. In the same sense that $E_J$ is the amplitude for a
pair-hopping between the leads and the grain, $J_+$ is the
amplitude for a Cooper-pair to tunnel between the two leads, skipping
over the grain (see Fig. \ref{weakfig1}). To be more specific, the
Cooper-pair hopping conductivities are
\begin{equation}
G_{AB}=G_{BC}\sim E_J^2,\hspace{5mm}G_{AC}\sim J_+^2.
\label{w1}
\end{equation}
 
Let us quote here the RG equations for the Josephson strengths
in the symmetric case\cite{Refael1}. To
distinguish between the bare Jospehson energy $E_J$ and the
renormalized one, we use $J$ to denote the flow of $E_J$. 
For the Josephson strengths $J$ of the junction and $J_+$ between the leads we get
\begin{eqnarray}
\frac{dJ}{dl}&=&J\l(1-\frac{R+r}{R_Q}\r)+\frac{R}{R_Q} J J_+,\label{w20a}\\
\frac{dJ_+}{dl}&=&J_+\l(1-\frac{2R}{R_Q}\r)+\frac{r}{R_Q} J^2.
 \label{w20b}
 \end{eqnarray}
The Josephson coupling $J_+$ between the leads is originally zero, but
will be generated under the flow in the second order in $J$. From these RG equations we can infer
the scaling behavior of the Cooper-pair conductivities in the
asymptotic low-temperature regime and in the ICFP area. This is done by
identifying the temperature $T$ with the RG scale as follows:
\begin{equation}
T\sim e^{-l}.
\label{rg_scale}
\end{equation}
\begin{figure}
\includegraphics[width=6cm]{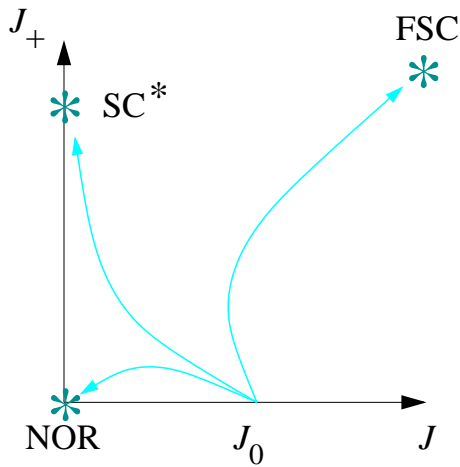}
\caption{Depending on the values of the resistors $r$ and $R$, the low
energy behavior of the two-junction system is described by three
stable  fixed points. In terms of the weak coupling Josephson
variables $J$ and $J_+$, these points are: (i) Normal (NOR)
fixed point at $J=J_+=0$, where Josephson tunneling is
suppressed. (ii) Fully Superconducting (FSC) point in which both $J$ and
$J_+$ are large and the flow is to strong coupling. (iii) \scs point
where $J_+$ flows to strong coupling and $J=0$. Here,
Cooper-pair tunneling is suppressed from lead to grain, but is very
strong between the two leads. \label{weakfig2}}
\end{figure} 

The normal phase of the system is described by the fixed point
$J=J_+=0$, in which the Josephson junctions are insulating (see Fig. \ref{weakfig2}). The Josephson coupling in this phase (and by Eq.~(\ref{w1}) thus also the
Cooper-pair conductivities) are expected to vanish as a power law in $T$,
\be
\ba{cc}
J\sim  T^{-\l(1-\frac{R+r}{R_Q}\r)}, & J_+\sim T^{-\l(1-\frac{2R}{R_Q}\r)}.
\ea
\label{w2}
\ee
The signature of the normal phase is a drop of both lead-to-lead and
lead-to-island conductance as temperature is reduced.  

If $J_+$ is relevant at first order ($R<R_Q/2$), then at low
temperatures points $A$
and $C$ in Fig.~\ref{two_junctions_refael} become short circuited,
so the leads become phase coherent {\it independently of $J$}. 
The RG equation for $J$ becomes
\be
\frac{dJ}{dl}=J\l(1-\frac{r+R/2}{R_Q}\r).
\label{w2d}
\ee
When $J$ is irrelevant ($r+R/2>R_Q$), the junctions are insulating
despite the phase coherence between the leads. This is the \scs
phase, and it is described by the fixed point shown in
Fig. \ref{weakfig2}. The conductivities of the individual junctions in
this phase vanish as the square of $J$, which follows the
power law
\be
J\sim  T^{\frac{r+R/2}{R_Q}-1}.
\label{w3}
\ee
The lead-to-lead conductance {\it diverges} as a power law of the
temperature and this divergence will be discussed in the next section using
the strong coupling analysis. Note that the signature of the \scs phase is
an increase in the lead-to-lead conductance, and a drop in the
lead-to-island conductance as the temperature decreases.

The FSC phase occurs when both $J$ and $J_+$ are relevant ($r+R/2<R_Q$). The weak
coupling equations describe the initial increase in the Josephson
couplings. The divergence of the conductance, however, can only be
described in the next section, where the strong coupling limit is discussed. 

\subsection{Strong coupling limit}

\begin{figure}
\includegraphics[width=8cm]{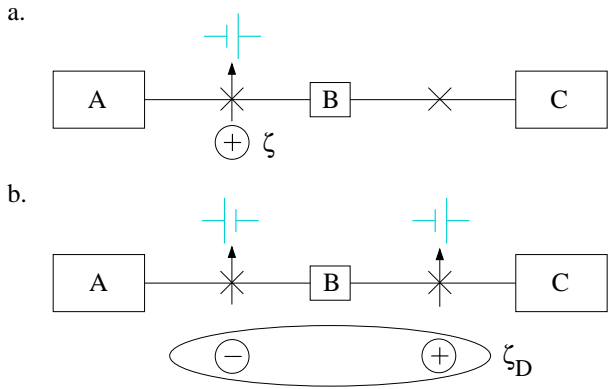}
\caption{In the strong-coupling limit we consider the RG flow of the phase slip fugacities $\zeta$ and $\zs$. (a) $\zeta$ is 
the amplitude, or fugacity, for a single phase-slip across either of the junctions that connect the leads to the middle grain. When a phase slip occurs it produces a momentary 
potential drop across the junction, which leads to dissipation. (b) When $r>0$, interactions 
between phase slips in the two junctions produce a phase-slip dipole in second order in the RG. The fugacity of the 
dipoles is $\zs$. When a dipole crosses the system there are two simultaneous momentary voltage drops with opposite directions on the two junctions, 
such that there is {\it no potential drop} between the leads, $A$ and $C$. \label{strongfig1}}
\end{figure}

When $E_J$ is large, one can no longer treat the Josephson coupling as a perturbation. But by employing
a duality, one can describe the system in terms of phase slips
\cite{Refael1}. Phase slips describe a sudden winding of a phase
difference across a junction. There are two kinds of phase slips we
need to consider: individual phase slips and phase-slip
dipoles. Individual phase slips create a potential drop, and hence
dissipation in a junction independently of the other junction; we
denote their {\it fugacity} by $\zeta$. A phase slip in junction $AB$
is partially screened if an anti phase-slip (a slip with the opposite
winding) simultaneously occurs in junction $BC$. This partial
screening leads to the generation of {\it phase-slip dipoles} in the
RG flow, in a very similar fashion to the generation of $J_+$. We denote the
dipole fugacity by $\zs$. By determining how often phase-slips
occur, $\zeta$ and $\zs$ determine the effective resistance across
the Josephson junctions (see Fig. \ref{strongfig1}):
\begin{equation}
R_{AB}=R_{BC}\sim \zeta^2+\zs^2, \hspace{5mm}  R_{AC}\sim 2\zeta^2.
\label{s1}
\end{equation}
The phase slip dipoles do not affect the lead-to-lead resistance since
a dipole produces a voltage blip in one junction and an opposite voltage blip in the other junction. Hence, the two blips add up
to zero, and do not produce a voltage drop between the
leads. 

The flow equations for $\zeta$ and $\zeta_D$ are:
\begin{eqnarray}
\frac{d\zeta}{dl}&=&\zeta\l(1-\frac{R_Q (R+r)}{2Rr+R^2}\r)+\frac{R R_Q}{R^2+2Rr} \zeta\zs,\label{s2a}\\
\frac{d \zs}{dl}&=&\zs\l(1-\frac{2R R_Q}{R^2+2Rr}\r)+\frac{r R_Q}{R^2+2Rr} \zeta^2.
\label{s2b}
\end{eqnarray}

\begin{figure}
\includegraphics[width=6cm]{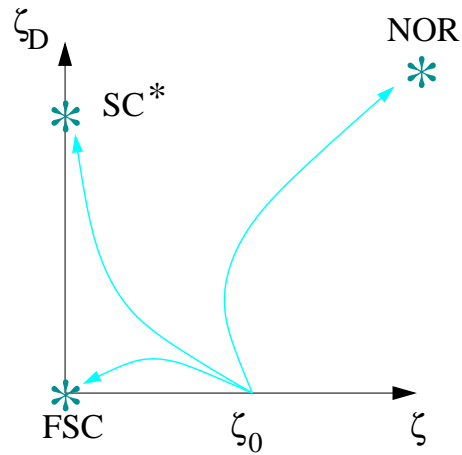}
\caption{In terms of the strong coupling phase-slip fugacities $\zeta$ and $\zs$, the three stable fixed  points are the following: (i) Fully superconducting  (FSC) fixed point at $\zeta=\zs=0$. In this point 
both single phase slips and phase-slip dipoles are suppressed. (ii) Normal (NOR) point in which both $\zeta$ and $\zs$ are large. There is no phase coherence and Cooper-pair tunneling in the system, which corresponds to weak coupling. (iii) The \scs point where
$\zs$ flows to strong coupling and $\zeta=0$. Phase coherence is maintained between the leads (points $A$ and $C$) but there is no phase coherence between the leads and the grain $B$. A quick look at Fig. \ref{weakfig2} reveals the strong-weak duality of this system.
\label{strongfig2}}
\end{figure}

In the FSC phase both $\zeta$ and $\zs$ are irrelevant, and decay to
zero as a power law in temperature (and so do the resistances by Eq.~(\ref{s1})),
\be
\ba{cc}
\zeta\sim T^{-\l(1-\frac{R_Q (R+r)}{2Rr+R^2}\r)}, & \zs\sim T^{-\l(1-\frac{2R R_Q}{R^2+2Rr}\r)}.
\ea
\label{s3}
\ee
The signature of the superconducting phase is thus a drop in both the
lead-to-lead and lead-to-grain resistance. This phase is controlled by
the $\zeta=\zs=0$ fixed point (Fig.~\ref{strongfig2}). Note that the FSC and
NOR phases are very similar in their behavior to the single
junction problem which has been investigated numerically in
Ref.~\onlinecite{junction}. 

When $\zs$ becomes relevant ($r+R/2>R_Q$), phase coherence is
destroyed between the leads and the central grain, but not between the two 
leads. At low temperatures, this practically implies that the effective resistance
of the resistor $r$ is diverging. Therefore the flow equation for
$\zeta$ becomes
\be
\frac{d\zeta}{dl}=\zeta\l(1-\frac{R_Q}{2R}\r).
\label{s4}
\ee
When $R<R_Q/2$, even though $\zs$ has diverged, $\zeta$ remains
irrelevant and decays to zero at low temperatures as
\be
\zeta\sim T^{\frac{R_Q}{2R}-1}.
\label{w2c}
\ee
In this phase the lead-to-lead resistance falls off as $\zeta^2$, but the 
lead-to-grain resistance diverges as $1/J^2$ from Eq.~(\ref{w3}). Again,
this is the \scs phase and the corresponding fixed point is shown in
Fig. \ref{strongfig2}. 

In the normal phase, both $\zeta$ and $\zs$ diverge at low energy scales, and one should use
Eqs.~(\ref{w20b}) and (\ref{w2}) to describe the low temperature behavior of
the measured resistances.

\subsection{Intermediate coupling fixed point (ICFP) region \label{ICFPth}}

\begin{figure}[t]
\centering
\includegraphics[angle=0, width=8cm]{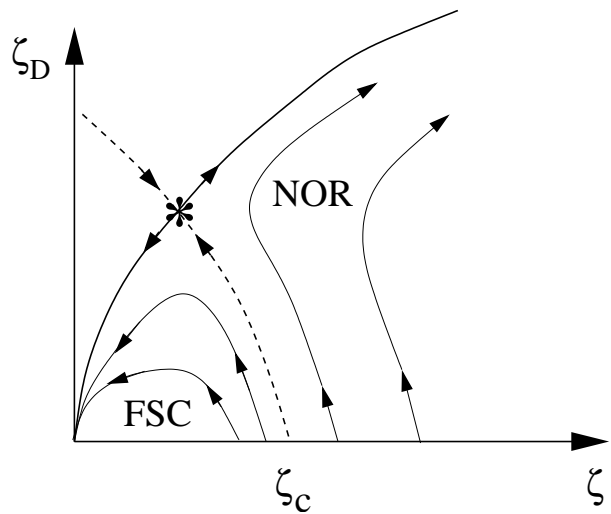}
\caption{Typical flow of the parameters $\zeta$ and $\zs$. If the initial
  values of $\zeta$ and $\zs$ are in the region marked FSC they flow
  towards $\zeta=\zs=0$ after flowing towards the intermediate
  coupling fixed point. If they are initially in the region marked
  NOR, they flow towards the strong coupling and the FSC phase, again,
  after flowing to the ICFP first. The dashed line is the critical
  manifold of the NOR-FSC transition, and the arrows near the
  fixed points mark the eigendirections of the linearized flow.  
 The physical line is $\zs=0$, and
  therefore the critical $\zeta_c$ for the NOR-FSC transition is given by
  the intersection of the critical manifold with the $\zeta_D=0$
  axis. }
\label{fig4}
\end{figure}

In the ICFP region a fourth fixed point appears (in addition to the
FSC, \scs, and NOR fixed points), and the scaling behavior of $J$ and $J_+$, or
$\zeta$ and $\zs$, is
determined at intermediate temperatures by this fixed point. At lower
temperatures the asymptotic $T\rightarrow 0$ scaling behavior is
determined by the $J=J_+=0$ (Normal phase) fixed point, or by the
$\zeta=\zs=0$ one (FSC phase). As opposed to the other regions of the phase diagram, where the resistors alone determine the phase, in the ICFP region, the phase of the system also depends on the
value of $E_J/E_C$. These parameters 
determine the critical value of the
initial amplitude of Cooper-pair hopping, or, using the strong
coupling picture, to the phase-slip fugacities.  In this region there
is no \scs phase, and the transition between the FSC and NOR phases is
direct. An illustration of this situation in the ICFP region is given in Fig.~\ref{fig4}. 

The existence of the unstable fixed point is evident from the
nonlinear flow equations for the weak and strong coupling
limits. We start with the weak coupling flow equations
(Eqs.~(\ref{w20a}) and (\ref{w20b})) written as follows
\begin{eqnarray}
\frac{dJ}{dl}&=&-J u+\frac{R}{R_Q} J J_+,\label{eq14a}\\
\frac{dJ_+}{dl}&=&-J_+ w+\frac{r}{R_Q} J^2,
 \label{eq14b}
\end{eqnarray}
where
\begin{equation}
u=\frac{R+r}{R_Q}-1, \hspace{5mm} w=\frac{2R}{R_Q}-1.
\label{eq14.5}
\end{equation}
When both $u>0$ and $w>0$, the RG
equations (\ref{eq14a}) and (\ref{eq14b}) have a third fixed point (in
addition to zero and
$\infty$). This point is at
\be
J^*=\frac{R_Q}{\sqrt{rR}}\sqrt{u w},\hspace{5mm}  J^*_+=\frac{R_Q}{R}u.
\label{eq15}
\ee
Therefore the lines $u=0$ and $w=0$ mark the weak-coupling boundaries of the
ICFP region 
(see Fig. \ref{pdtheory}).  

Similarly, in the strong coupling limit we can write the flow
equations (\ref{s2a}) and (\ref{s2b}) as
\begin{eqnarray}
\frac{d\zeta}{dl}&=&-\zeta\overline{u}+\frac{R R_Q}{R^2+2Rr} \zeta\zs,\\
\frac{d \zs}{dl}&=&-\zs\overline{v}+\frac{r R_Q}{R^2+2Rr} \zeta^2,
\label{i16}
\end{eqnarray}
where
\begin{equation}
\overline{u}=\frac{R_Q (R+r)}{2Rr+R^2}-1,\hspace{5mm} \overline{w}=\frac{2R R_Q}{R^2+2Rr}-1.
\label{i17}
\end{equation}
As in the weak coupling limit, when $\overline{u}>0$ and
$\overline{w}>0$, an unstable fixed point appears at intermediate
values of $\zeta$ and $\zs$. The lines  $\overline{u}>0$ and
$\overline{w}>0$ mark the boundaries of the ICFP region on the strong
coupling side. The fixed point fugacities of the ICFP are given by
\be
\zeta^*=\frac{R^2+2rR}{R_Q \sqrt{rR}}\sqrt{\overline{u}\overline{w}},\hspace{5mm}
\zs^*=\frac{R^2+2rR}{R_Q R}\overline{u}.
\label{i18}
\ee
The scaling properties of the system near criticality and at
intermediate energy scales are determined by the critical properties
of the above ICFP and, in particular, the relevant and irrelevant
directions and the exponents associated with them: $\lambda_+$ and
$\lambda_-$. These properties are calculated in Appendix \ref{appA}. 

Qualitatively the situation is illustrated in Fig. \ref{fig4}. 
Consider a generic flow of $\zeta$ and $\zs$ at some value of the resistors $r$
and $R$. If the system we are considering has initial fugacities $\zeta^{(0)}$
and $\zs^{(0)}$ in the FSC-region, then both $\zeta$ and $\zs$ flow towards 0 eventually, and
the system is in the superconducting phase at low energies. But this
behavior is not at all trivial: before decaying to zero, $\zs$ starts
off growing as $\zeta$ decreases. Similarly, if $\zeta^{(0)}$ and
$\zs^{(0)}\sim 0$ are in the NOR region, once the flow passes by the ICFP, both $\zeta$ and
$\zs$ become relevant. But before the flow reaches the fixed point,
there is an energy range in which $\zs$ grows, but $\zeta$
decreases. Generally speaking, the extent of this range of energies is determined by
the irrelevant critical exponent, $\lambda_-$ of the ICFP (see Appendix \ref{appA}). 

The FSC-NOR critical point occurs when $\zeta^{(0)}$
and $\zs^{(0)}$ are on the critical manifold which flows exactly into
the ICFP. In the two junction system $\zs^{(0)}=0$, and therefore
there is a critical fugacity $\zeta_c$ above which the system is in the
NOR phase, and below which it is in the FSC phase. The initial phase
slip fugacity is determined by the ratio $E_J/E_C$, and therefore this
transition (for a given value of the resistances $R$ and $r$ in the
ICFP region) can be tuned by changing the {\it Josephson energy}. The
qualitative picture described above is equally valid in weak coupling,
where the only difference would be discussing the pair-tunneling
amplitudes, $J$ and $J_+$, rather than the phase-slip fugacities.

The flow of the fugacities (or pair-tunnel amplitudes) before reaching the ICFP
determines the behavior of the resistance as a function of temperature
at intermediate temperatures. In these temperature ranges the behavior
of the resistance may be misinterpreted as any of the three phases of
the system. Particularly, in the region of parameter space where
$\zeta$ decreases  and $\zs$ increases, the lead-to-lead resistance
decreases, since it is only proportional to $\zeta^2$, but the
lead-to-grain resistance increases. This behavior could be
misinterpreted as the system being in the \scs phase.
In order to determine the true $T=0$ phase one has to
investigate the system at very low temperatures. These crossover
effects indeed appear explicitly in the Monte Carlo simulations of the system.

\section{Monte Carlo Method \lb{mcm}}

Monte Carlo simulations of the two-junction model as shown in the lower part of Fig.~\ref{two_junctions_refael} can be performed using variants of the local and cluster updates detailed in Ref.~\onlinecite{junction}. Imaginary time is divided into $N$ time slices of size $\Delta \tau = \frac{\beta}{N}$ and the variables $\phi_{1,n}$ and $\phi_{2,n}$, $n=1, \ldots , N$ (phase differences across the junctions at the discrete times $\tau_n=n\Delta\tau$) are used to represent the phase configuration. We implemented the following types of Monte Carlo updates:

\begin{enumerate}
\item Single junction cluster updates
  
A cluster of connected sites is constructed in one of the junctions as outlined in Ref.~\onlinecite{junction}. The cost in action of flipping the cluster, $\Delta S_\text{lead}$, which is associated with the last term in Eq.~(\ref{S_D}),
\begin{eqnarray}
S_\text{lead}&=&\frac{R_Q}{R_\text{lead}}\int_0^\beta d\tau d\tau' \frac{(\pi/\beta)^2}{\sin^2((\pi/\beta)(\tau-\tau'))}\nonumber\\
&&\times((\phi_1(\tau)+\phi_2(\tau))-(\phi_1(\tau')+\phi_2(\tau')))^2,\nonumber\\
\label{s_lead}
\end{eqnarray}
must be calculated and the cluster move accepted with probability
\begin{equation}
p = \min(1, \exp(-\Delta S_\text{lead})).
\end{equation}

\item Single junction local updates 

Local updates in Fourier-space are proposed in one of the junction, as detailed in Ref.~\onlinecite{junction}. They, too, are accepted with probability
\begin{equation}
p = \min(1, \exp(-\Delta S_\text{lead})).
\end{equation}  

\item Compensated single junction cluster updates

The SC$^*$-phase is characterized by phase coherence between the leads ($\phi_1(\tau)+\phi_2(\tau)\approx \text{const}$) but strong fluctuations of the variables $\phi_1(\tau)$ and $\phi_2(\tau)$ (insulating junctions). Therefore, the fluctuations in the two junctions essentially compensate each other and efficient updates in the SC$^*$-phase should take this constraint into account. 
 
 In a compensated single junction cluster update, a cluster of connected sites is constructed in one of the junctions as outlined in Ref.~\onlinecite{junction} and an equal, but opposite move proposed for each phase variable in the other junction. Since $\Delta \phi_1+\Delta \phi_2 = 0$, there is no cost in action associated with the $S_\text{lead}$-term in Eq.~(\ref{S_D}). However, there is a cost in action, $\Delta S_\text{oj}$, associated with the contribution from 
 the other junction (hence the notation ``oj"),
\begin{eqnarray}
S_\text{oj}&=& \frac{1}{16E_C}\int_0^\beta d\tau \Big( \frac{d\phi_\text{oj}}{d\tau}\Big)^2-E_J \int_0^\beta d\tau \cos(\phi_\text{oj})\nonumber\\
&&+\frac{R_Q}{R_\text{junction}}\int_0^\beta d\tau d\tau' \frac{(\pi/\beta)^2(\phi_\text{oj}(\tau)-\phi_\text{oj}(\tau'))^2}{\sin^2((\pi/\beta)(\tau-\tau'))}.\nonumber\\
\end{eqnarray}
The compensated cluster move should therefore be accepted with probability
\begin{equation}
p = \min(1, \exp(-\Delta S_\text{oj})).
\end{equation} 

\begin{figure}[t]
\centering
\includegraphics[angle=0, width=8cm]{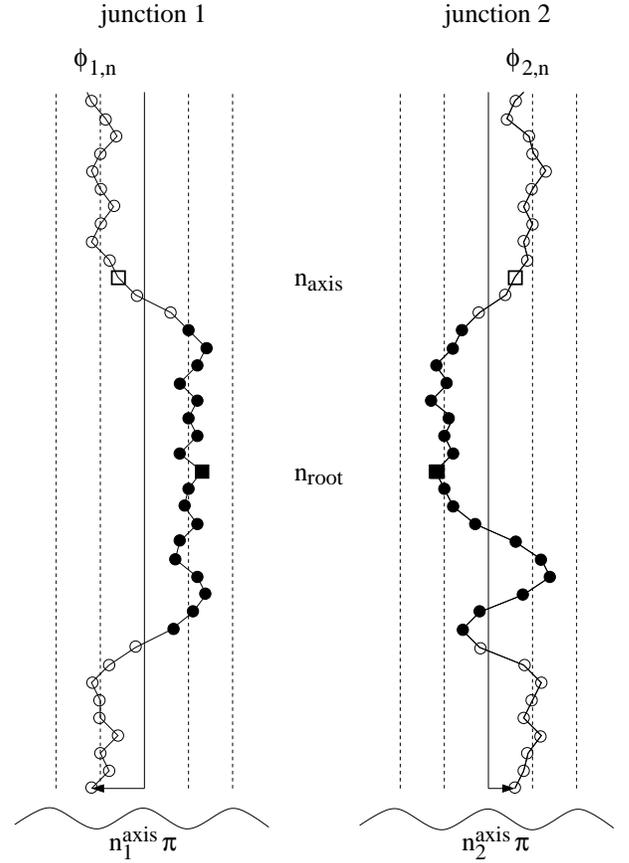}
\caption{Illustration of the two-junction cluster updates. The axis is
  chosen in at a symmetry point of the potential neighboring
  $\phi_{i,n_\text{axis}}$ ($n_\text{axis}$ is a random site), but on
  opposite sides in the junction $i=1$ and 2. A second random site
  $n_\text{root}$ is picked as the root site of the cluster. The sites
  connected to the root site are determined using the bond
  probabilities (\ref{double_cluster1}) and are marked with black
  dots. The new configuration is obtained by flipping the cluster
  around the axis. In other words, the spin like variables defined
  relative to the axis and shown as an arrow on the first site are
  inverted during a cluster update.}
\label{fig_mc}
\end{figure}

\item Two-junction cluster updates

A random site $k$ is picked and an axis $n_i$ in each of the two junctions chosen among the two closest to $\phi_i(\tau_k)$, such that $n_1\pi\le \phi_{1,k}$ and $n_2\pi\ge \phi_{2,k}$ or vice versa. Relative coordinates $\phi^\text{axis}_i=\phi_i-n^\text{axis}_i\pi$ are introduced in both junctions and a cluster of sites connected to the root-site $k$ is constructed using the bond probabilities
\begin{eqnarray}
p(k,l) &=& \max(0, 1-\exp(-\Delta S_{k,l}))
\label{double_cluster1}
\end{eqnarray}
where the cost in action of breaking a bond, $\Delta S_{k,l}$, is defined as
\begin{eqnarray}
\Delta S_{k,l} &=& \sum_{i=1,2} \Big( S(\phi^\text{axis}_{i,k},-\phi^\text{axis}_{i,l})- S(\phi^\text{axis}_{i,k},\phi^\text{axis}_{i,l})\Big)\nonumber\\
&=& 8g(k-l)\sum_{i=1,2} \phi^\text{axis}_{i,k}\phi^\text{axis}_{i,l}.
\label{double_cluster2}
\end{eqnarray}
In the above expression, $g(j)$ is the kernel ($j \ne 0$) 
\begin{eqnarray}
g(j) &=&  \frac{1}{32E_C\Delta\tau}(\delta_{j,1}+\delta_{j,N-1})\nonumber\\
&+&\frac{1}{8\pi
^2}\frac{R_Q}{R_\text{junction}}\frac{(\pi/N)^2}{\sin((\pi/N)j)^2}.\label{g}
\end{eqnarray}
Hence, in contrast to the single junction case, a cluster can contain relative phase variables of both signs. The cluster building process (\ref{double_cluster2}) takes into account the capacitive, dissipative and Josephson contributions from both junctions, but not the dissipative contribution from the $S_\text{lead}$-term. A two-junction cluster move therefore can only be accepted with probability 
\begin{equation}
p = \min(1, \exp(-\Delta S_\text{lead})).
\end{equation}  
  
\end{enumerate}


\begin{figure}[t]
\centering
\includegraphics[angle=-90, width=8.5cm]{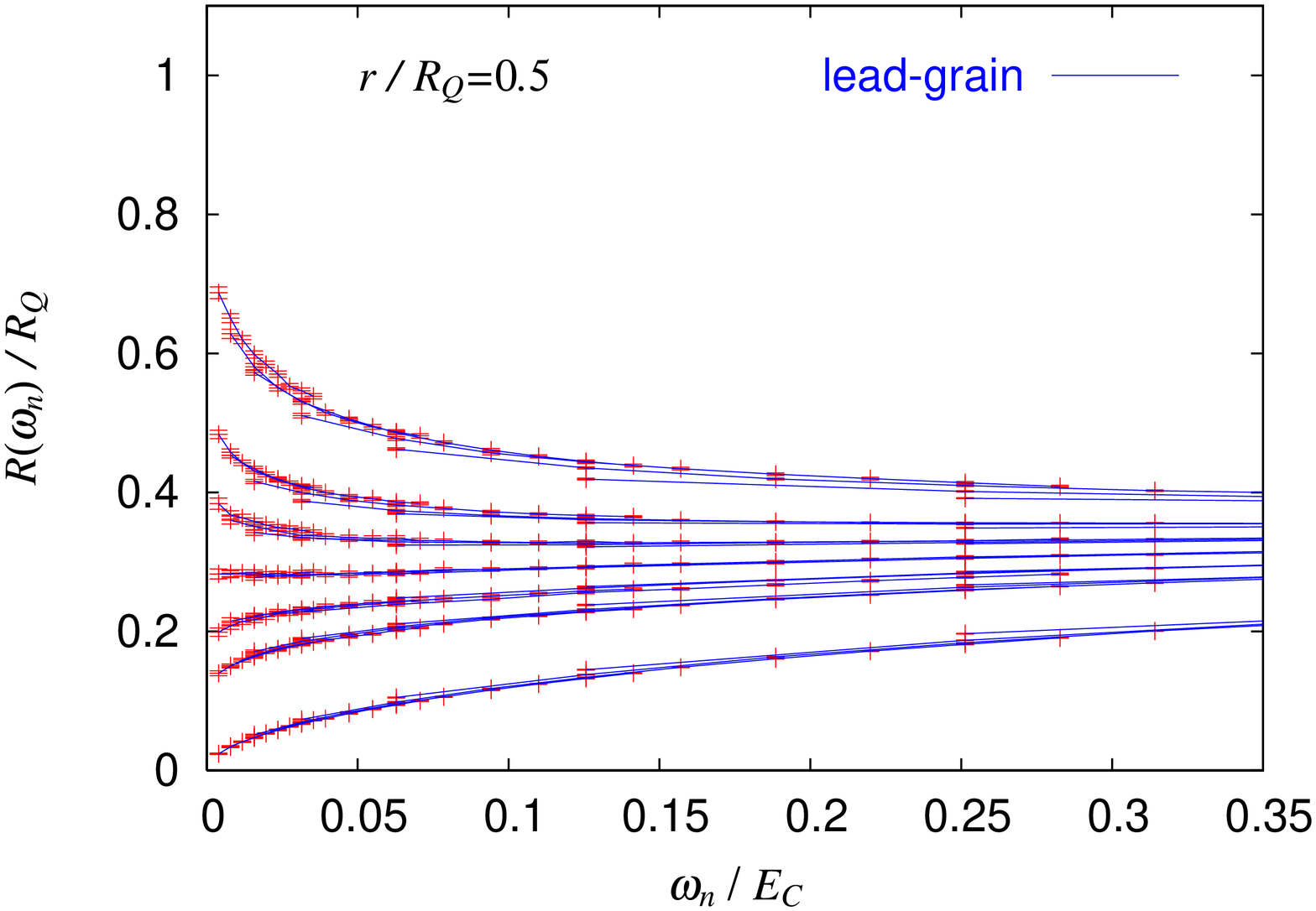}
\includegraphics[angle=-90, width=8.5cm]{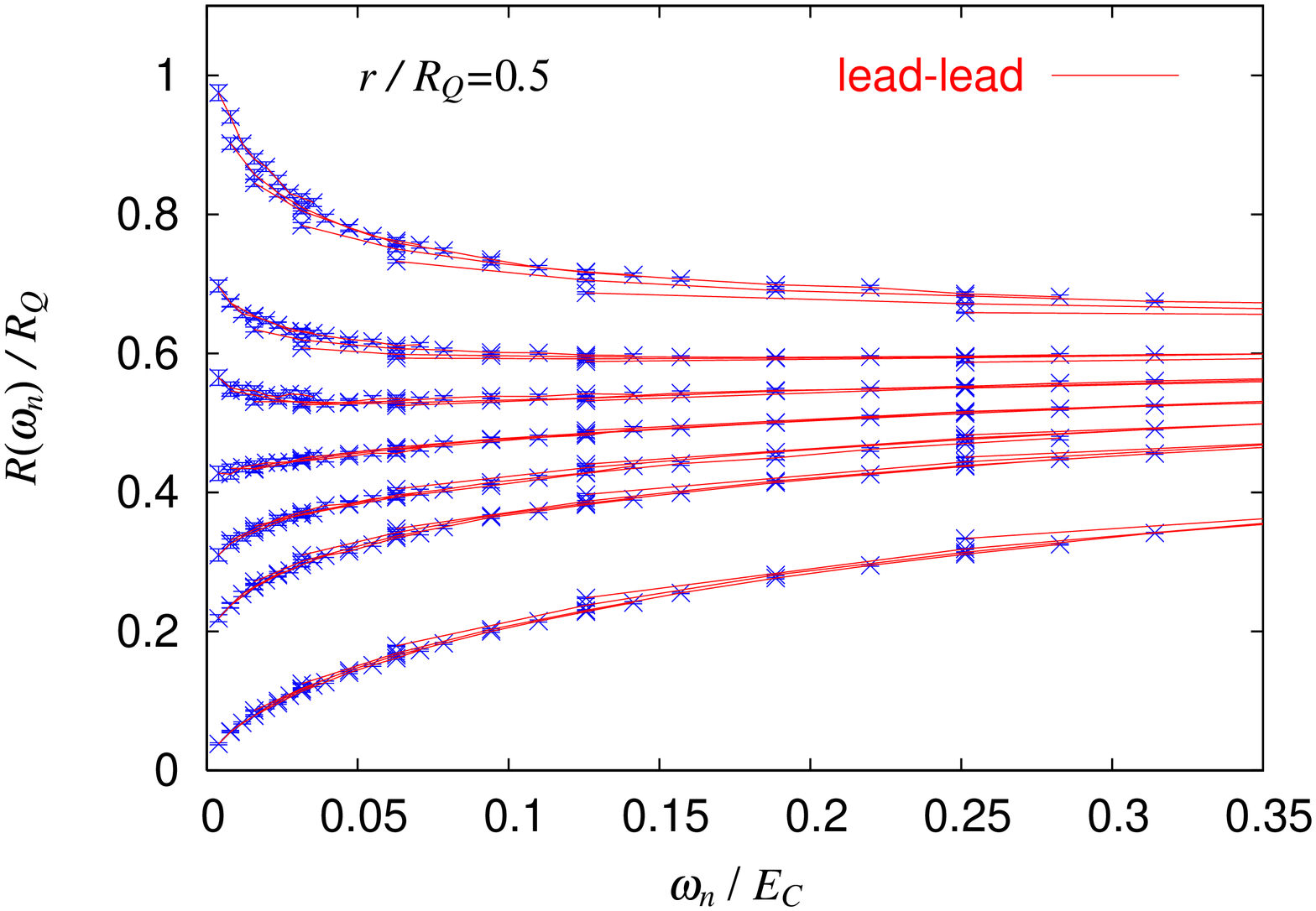}
\caption{Imaginary frequency resistances for $r=0.5$ and $\beta E_C=25$, 50, 100, 200, 400, 800 and 1600. Only the first 10 Matsubara points are shown for each temperature and the extrapolation of these curves to zero frequency gives the zero-bias resistance. From bottom to top, the different sets of curves correspond to $R=0.5$, 0.6, 0.625, 0.65, 0.675, 0.7 and 0.75. The curves in the upper panel show the resistance from lead to central grain, those in the lower panel the resistance from lead to lead. In both cases the superconductor-to-normal phase transition occurs at $R\approx 0.65$.}
\label{sc_n}
\end{figure}
\begin{figure}[t]
\centering
\includegraphics[angle=-90, width=8.5cm]{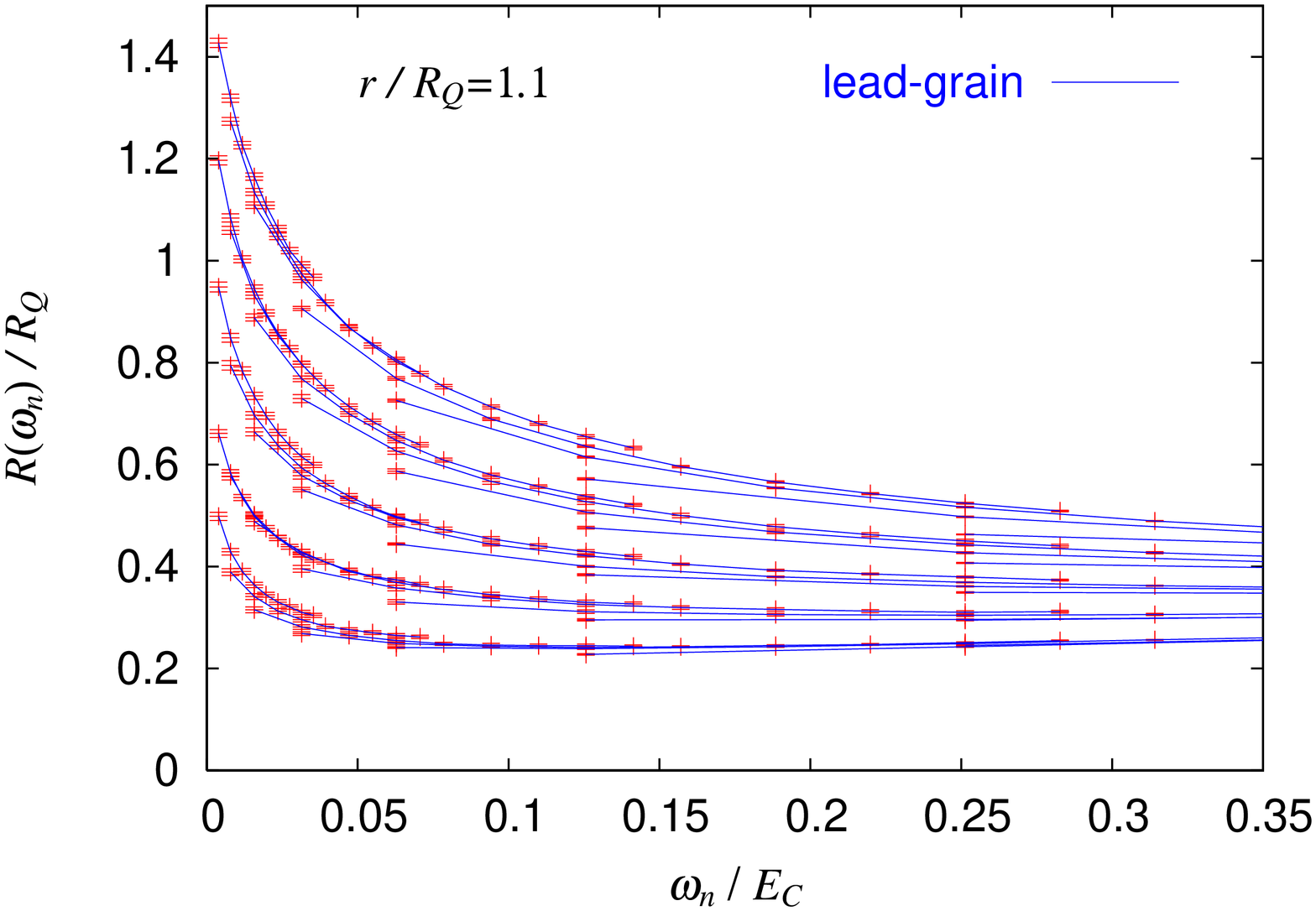}
\includegraphics[angle=-90, width=8.5cm]{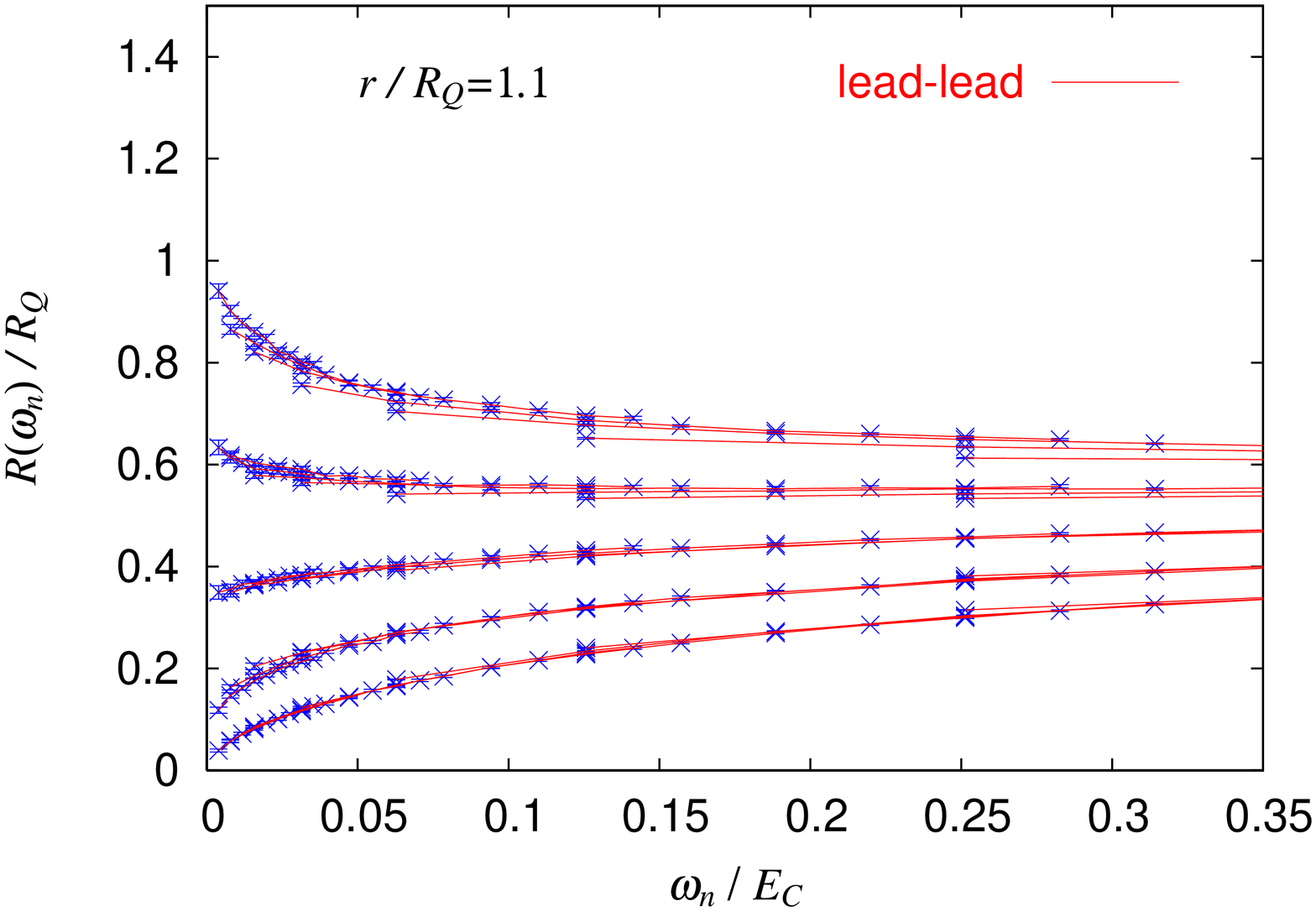}
\caption{Imaginary frequency resistances for $r=1.1$ and $\beta
  E_C=25$, 50, 100, 200, 400, 800 and 1600. Only the first 10 Matsubara points are shown for each temperature and the extrapolation of these curves to zero frequency gives the zero-bias resistance. From bottom to top, the different sets of curves correspond to $R=0.4$, 0.45, 0.5, 0.55
  and 0.6. The curves in the upper panel show the resistance from
  lead to central grain, those in the lower panel the resistance
  from lead to lead. For $R<0.5$, the junctions are insulating, but
  the device from lead to lead is superconducting (SC$^*$ phase).}
\label{sc*_n}
\end{figure}

\section{Phase diagram}

We first use the efficient Monte Carlo scheme outlined in
Sec. \ref{mcm} to identify the
three phases NOR, FSC and \scs, and 
to determine the phase diagram for an intermediate value of the Josephson
coupling $E_J$. This allows to test the theoretical predictions 
 outlined in Sec.~\ref{theory} and in Fig.~\ref{pdtheory}. 

The result of this study is shown in Fig. \ref{phasediagram} and
explained in Sec. \ref{simres}. The agreement between the Monte Carlo calculation and the theory is
very good.  Small deviations, however, appear in the vicinity of the
(tricritical) meeting point of the three phases: $r=0.75R_Q$ and $R=R_Q/2$. These deviations from the theoretically predicted
phase diagram are explained in Sec. \ref{COsec} as crossover effects, and
are indirect evidence for the existence of the intermediate coupling fixed point.


\subsection{Simulation results \lb{simres}}

We use the resistance at imaginary
frequencies to identify the state of conductance between the leads and
from the leads to the central grain. The imaginary frequency
resistance is defined as
\begin{equation}
\frac{R(\omega_n)}{R_Q} = \frac{1}{2\pi}|\omega_n|\langle \phi\phi \rangle_{\omega_n},
\end{equation}
where $\omega_n=(2\pi n)/\beta$ denotes a Matsubara frequency and
$\langle \phi\phi \rangle_{\omega_n}$ the Fourier transform of the
phase-phase correlation function $\langle \phi(0)\phi(\tau) \rangle$ ($\phi\equiv \phi_i$ in the case of
conductance from lead to central grain and $\phi\equiv \phi_1 +
\phi_2$ for the conductance from lead to lead).

In Figs.~\ref{sc_n} and \ref{sc*_n} we show the results of such an
analysis obtained for $E_J/E_C=1$ and $\Delta\tau E_C = 0.25$. The
interpretation of the data is the same as in the single-junction case
discussed in Ref.~\onlinecite{junction}. Figure~\ref{sc_n} shows the
transition across the FSC-NOR phase boundary at $r=0.5R_Q$. The different
sets of curves correspond to different values of $R$ and the different
lines in each set of curves to different temperatures. Only
the lowest ten Matsubara frequencies are shown. Extrapolating these
curves to $\omega_n\rightarrow 0$ yields the zero bias resistance,
which decreases to zero with decreasing temperature in the $T=0$
superconducting state. If the junction turns insulating with
decreasing temperature, the resistance increases and eventually
saturates at the value $r+R$ (from lead to central grain) or $2R$
(from lead to lead). It is obvious from the data in Fig.~\ref{sc_n},
that the transition occurs simultaneously in the individual junctions
and between the leads, in accordance with theoretical predictions. 

\begin{figure}[t]
\centering
\includegraphics[angle=-90, width=8.5cm]{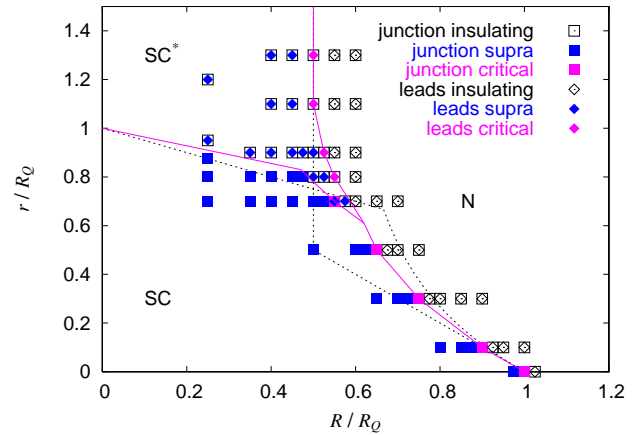}
\caption{Phase diagram obtained from the analysis of the
  resistance-versus-temperature behavior. We find a good agreement
  with theoretical predictions (black dotted lines), except in the
  region around the tricritical point, where the three phases meet.}
\label{phasediagram}
\end{figure}

\begin{figure}[t]
\includegraphics[angle=-90,width=8cm]{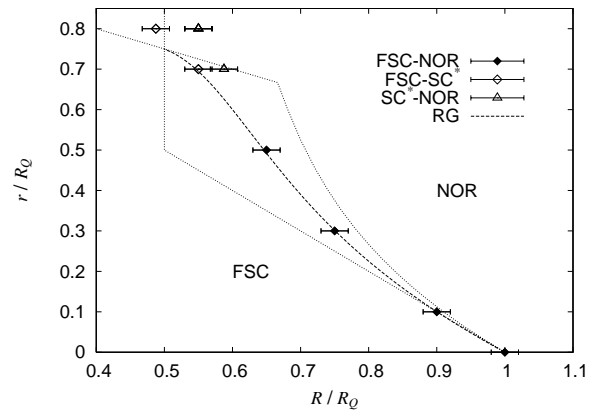}
\caption{A closeup of the phase diagram in the ICFP
  region for $E_J/E_C=1$. The dashed line is the RG prediction for the FSC-NOR phase
  boundary as obtained in App. \ref{appB}, with the bare phase slip
  fugacity being the only fit parameter, determined using the
  critical point at $r=R_Q/2$ to be $\zeta_0=0.295$. The
  RG-results agree well with the Monte Carlo results. 
 Filled dots mark the FSC-NOR phase boundary and empty dots a FSC-\scs or \scs-FSC transition. 
\lb{close-up}
  }
\end{figure}

Figure \ref{sc*_n} shows the transition across the SC$^*$-NOR phase
boundary at r=1.1. First of all, we note that the data for $R<0.5R_Q$
clearly prove the existence of the SC$^*$ phase. While the individual
junctions turn insulating as $T\rightarrow 0$, the resistance from lead to
lead decreases to zero as the temperature decreases ($T=0$
superconductivity). If $R$ is increased, however, the device undergoes
a (lead to lead) superconductor-to-metal transition at $R=0.5R_Q$, as
predicted by theory.

Repeating this type of analysis for several values of $r$ we could map
out the phase diagram which is shown for $E_J/E_C=1$ in
Fig.~\ref{phasediagram}. As expected, the FSC-NOR phase boundary lies
somewhere in between the limiting values calculated analytically for
$E_J/E_C \gg 1$ and $E_J/E_C \ll 1$, indicated by the dotted lines
(see also Fig.~\ref{phasediagram_refael}). A close-up of the FSC-NOR
phase boundary is shown in Fig. \ref{close-up}; the measured critical
line at $E_J/E_C=1$ agrees very well with an RG-based calculation
(presented in App. \ref{appB}).

Overall, we find a good
agreement with the theoretically predicted phase diagram, except
in the vicinity of the tricritical point, $r=0.75R_Q,\,R=R_Q/2$, where
the three phases meet. Phase boundaries there appear to be shifted to a
somewhat larger value of $R$ and a smaller value of $r$. As it turns
out, in this
region, crossover effects are important and determining the phase from
finite-temperature simulations can be misleading. A detailed
discussion of the observed deviations is given below, in Sec. \ref{COsec}.

\begin{figure}[t]
\centering
\includegraphics[angle=-90, width=8.5cm]{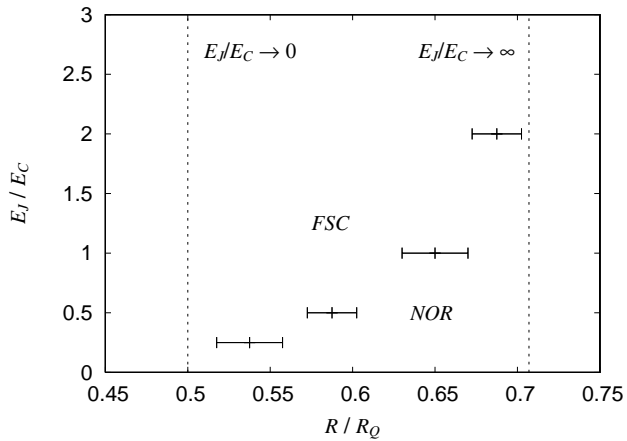}
\caption{Phase boundary as a function of the Josephson coupling strength calculated at $r/R_Q=0.5$. The dashed lines indicate the analytically predicted values for the critical $R$ in the weak- and strong-coupling limits.}
\label{boundary_E_J}
\end{figure}

As the Josephson coupling strength $E_J$ is increased from zero to
infinity, the FSC-NOR phase boundary shifts from the weak coupling to the
strong coupling limit indicated in Fig.~\ref{phasediagram_refael}. This
behavior is demonstrated in Fig.~\ref{boundary_E_J}, which shows 
a cut across the critical surface at $r/R_Q=0.5$.  

\subsection{Crossover effects in the Monte Carlo results \lb{COsec}}

The theory of the two-junction system as outlined in Sec.~\ref{theory}
and in App.~\ref{appA} leads us to expect crossover behavior at
intermediate temperatures. Although the range of temperatures  at our
disposal is limited, and the crossover regime spans a narrow range of
energy scales, we see indirect evidence of the crossover effects in
the measured phase diagram.  As mentioned above, near the meeting
point of the three phases, the predicted phase boundaries do not
completely agree with the Monte Carlo simulation results (Fig. \ref{phasediagram}). We will explain these deviations using the RG analysis.

The deviations of the measured data from the theoretical
predictions can be understood by looking at the RG-trajectories given
by Eqs.~(\ref{s2a}) and (\ref{s2b}). As explained in Sec.~\ref{ICFPth}, the flow of the
phase slip fugacities or pair-tunneling amplitudes near the
intermediate coupling fixed point can make the
measured resistance as a function of temperature behave as though the
system is in the \scs phase. Near the meeting point of the three phases, the flow towards the ICFP
is extremely slow; it is dominated by the critical exponent
$\lambda_-$ given by Eqs.~(\ref{eq19a}) and (\ref{eq19sa}):
\begin{eqnarray}
\lambda_-^{(\text{weak})}&=&\frac{1}{2}\l(-w-\sqrt{w^2+8uw}\r),\label{10a}\\
\lambda_-^{(\text{strong})}&=&\frac{1}{2}\l(-\overline{w}-\sqrt{\overline{w}^2+8\overline{u}\overline{w}}\r),
\label{10b}
\end{eqnarray}
where $u,\,w$ and $\overline{u},\,\overline{w}$ are defined in
Eqs.~(\ref{eq14.5}) and (\ref{i17}).  
Loosely speaking, but more specifically, in this regime
$x-x^{*}\sim T^{\lambda_-}$ with $x$ being $\zeta$, $\zs$, or $J$,
$J_+$, and $x^{*}$ the respective fixed-point value.  As can be seen
from Eqs.~(\ref{10a}),  (\ref{10b}) and Fig.~\ref{pdtheory}, $\lambda_-$ in both the
weak and  strong coupling regimes vanishes at the meeting point
($w=\overline{w}=0$), and
therefore is expected to be small near this special point for any
value of the Josephson coupling.  This implies that since the
Monte Carlo calculation is limited to temperatures above $T=E_c/2500$
(if we choose $\Delta \tau E_C = 0.25$, which seems appropriate)
it may not probe the ground state of the system, but rather the
crossover physics. This will lead to distorted NOR-FSC phase
boundaries in this region. As it turns out, this phenomenon is not
restricted to  the ICFP region: slow crossovers occur all around the
triple point and may shift the observed NOR-\scs and \scs-FSC phase
boundaries, as indeed is seen in Fig. \ref{phasediagram}.

\begin{figure}
\includegraphics[width=7.5cm]{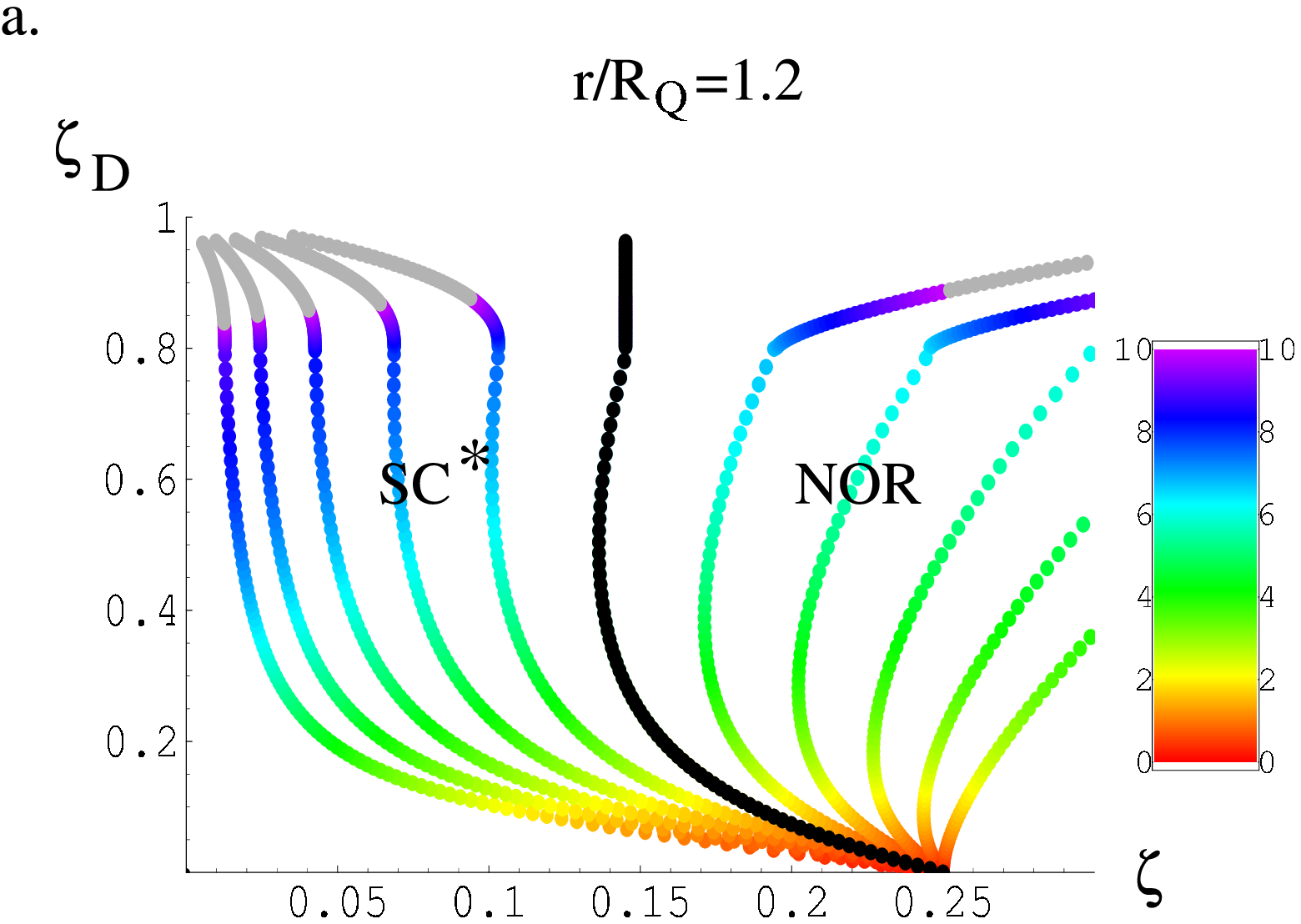}
\includegraphics[width=7.5cm]{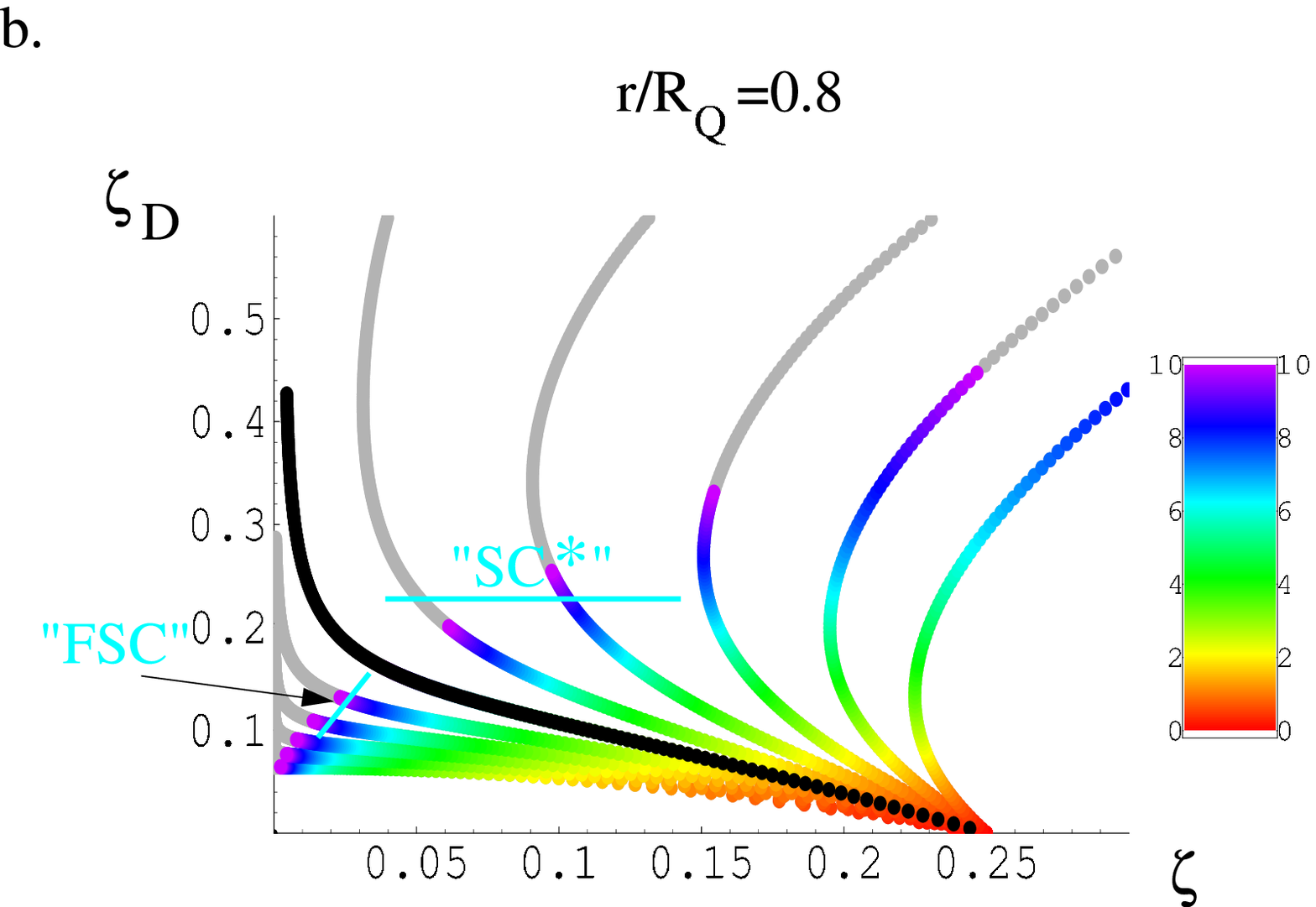}
\caption{Simulated RG flows of phase slip fugacities (Eqs.~(\ref{s2a}) and (\ref{s2b})). The initial
  conditions for all plots are $\zeta^{(0)}=0.25$, $\zs^{(0)}=0$. The color-coding
  marks the RG flow parameter $l=\ln \Lambda/T$, and 
  regions not accessible with
  Monte Carlo simulations ($l>10$) are shown in light gray. (a)
  $r/R_Q=1.2$, $R/R_Q$ varies from $0.4$ to $0.6$ from left to
  right. The black curve is at the critical value $R=R_Q/2$. Within the 
  energy scales probed by the Monte Carlo simulation it is easy to
  distinguish between the \scs and NOR phases. In the \scs phase
  $\zeta$ decreases making the lead-to-lead resistance $R_{AC}\sim
  \zeta^2$ decrease at low energies, while $\zs$ grows rapidly,
  producing an insulating regime for the effective resistance for each
  junction, $R_{AB}$. We assume that $\zs$ begins to saturate at
  $\zs=0.7$ and makes the flow cross over to Eq.~(\ref{s4}). (b)
  $r/R_Q=0.8$, and $R/R_Q$ varies as in (a). At this value of $r$ the
  temperature range accessible with simulations is no longer sufficient to determine the
  true phase boundaries. For a range of $R$ values near and above the
  critical point $R=R_Q/2$ we may mistakenly identify the phase as \scs,
   since $R_{AC}\sim \zeta^2$ is decreasing, but $R_{AB}\sim
  \zeta^2+\zs^2$ is increasing. Similarly, for $R<R_Q/2$ it looks as
  if the system is in the FSC phase (instead of the \scs phase), since
  $\zs$ grows very slowly. This explains the deviations of the numerically determined phase
  boundaries (Fig.~\ref{phasediagram}) from
  the theoretical ones.
\label{rfiga}
\label{cflowa}}
\end{figure}

\begin{figure}
\centering
\includegraphics[width=7.5cm]{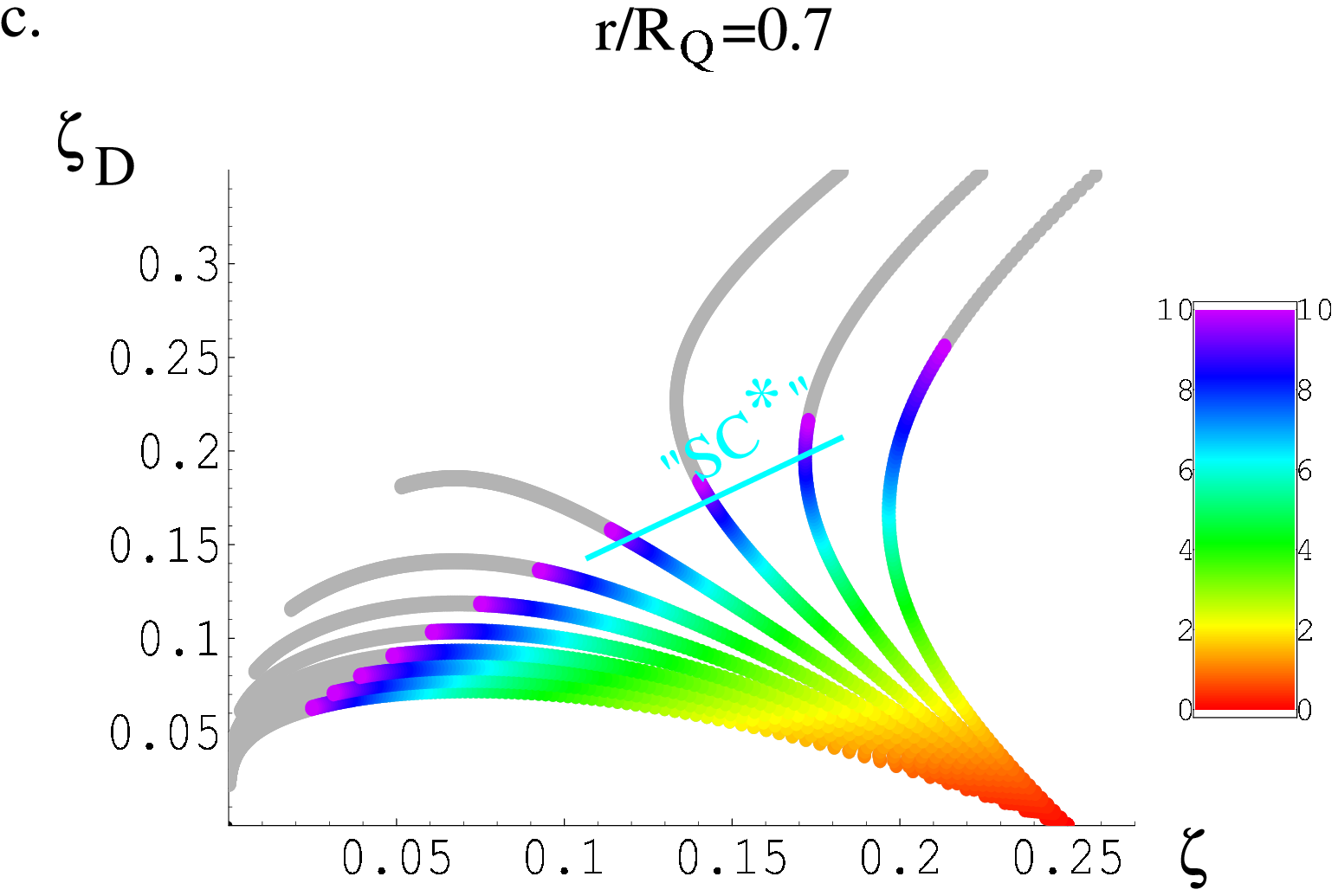}
\includegraphics[width=7.5cm]{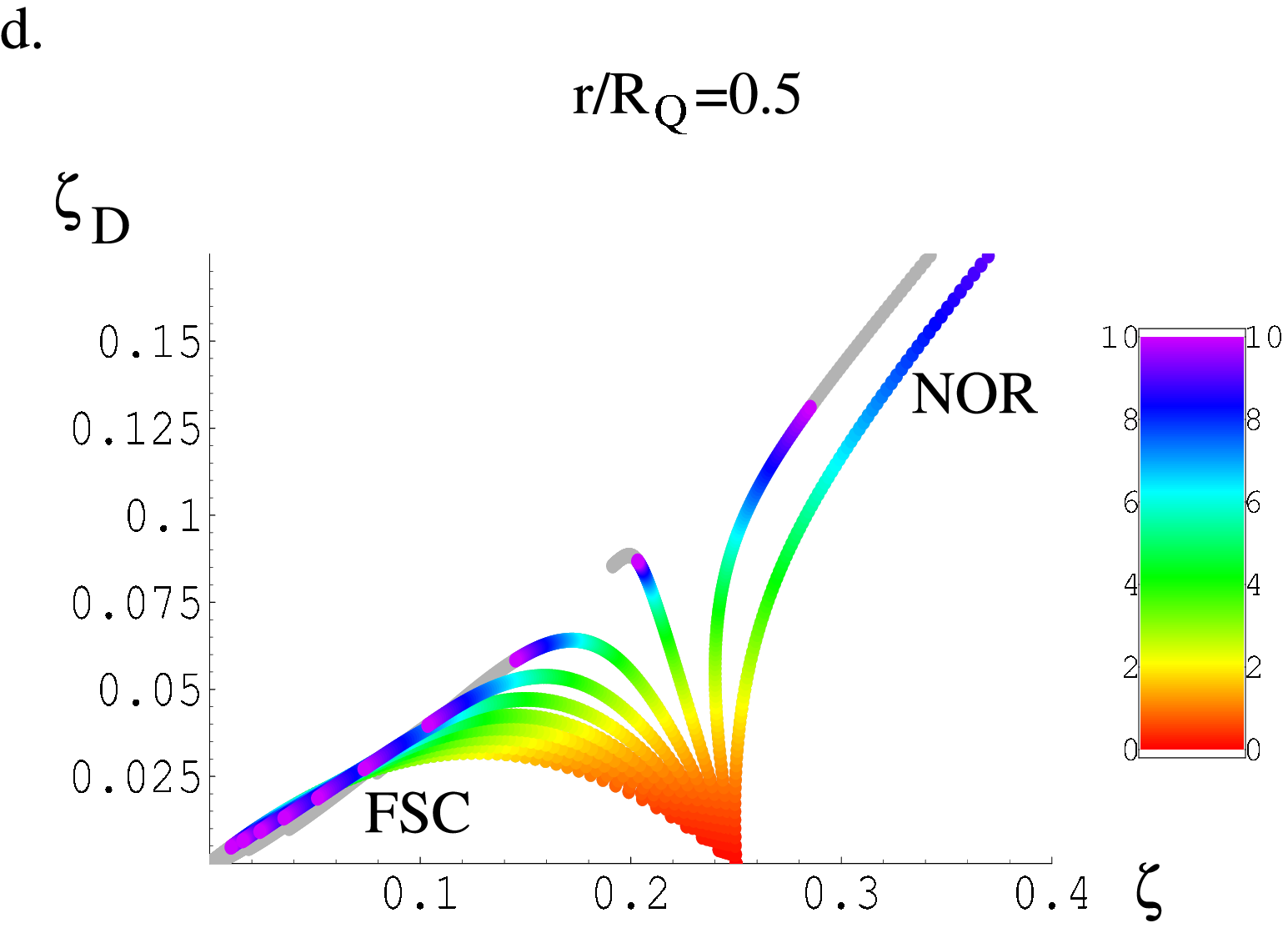}
\caption{
Same as Fig.~\ref{rfiga}:
(c) $r/R_Q=0.7$, $R/R_Q$ varies between $0.5$ and $0.6$ from right to
  left. These flows are in the ICFP region. Even though in this
  region we expect a direct NOR-FSC transition, at the energy range
  accessible in the Monte Carlo simulation we again see a region which would be
  mistaken as \scs, which is indeed the case in
  Fig. \ref{phasediagram}. The slow crossover is due to the proximity
  to the triple point at $r=0.75R_Q$ where the three phases meet.
(d) $r/R_Q=0.5$, and $R/R_Q$ varies as in (c). As in Fig. \ref{rfiga}a even within the energy range accessible
  numerically, we clearly see the NOR and FSC phases showing up --
  $\zeta$ and $\zs$ either both diverge or both decrease at the
  lowest temperatures accessible by Monte Carlo simulation. 
\label{rfigb}
\label{cflowb}}
\end{figure}

To demonstrate the above behavior we plotted the predicted RG flow of the phase slip fugacities in the 
strong coupling limit for four values of $r/R_Q$ in the region of interest (Figs. \ref{cflowa} and \ref{cflowb}). We color coded the plot 
according to the RG flow parameter $l=\ln \Lambda/T$ where $\Lambda$ is
the ultra-violet cutoff. Roughly speaking, the maximum RG scale we probe in the
Monte Carlo calculation is $l_{\text{max}}\approx 7$, and therefore we stop the color coding at the
value $l=10$. The various flow lines all start with
the same initial fugacity (same $E_J/E_C$), but have the 
resistance $R$ varied across the transition. In the Monte Carlo simulation one measures the lead-to-lead resistance, which we expect to behave as
\be
R_{AC}\sim \zeta^2,
\ee
and the lead-to-grain resistance, which goes as
\be
R_{AB}=R_{BC}\sim \zeta^2+\zs^2.
\ee
As can be seen from Fig.~\ref{cflowa}a, at large values of $r$, $r/R_Q=1.2$, the flow within 
the accessible range of $l$ allows an easy determination of the phase: at low energies, $\zeta$ slowly decays or diverges to the left or the right of the  
black critical flow line at $R=R_Q/2$, while $\zs$ strongly diverges
(for this illustration we assumed that when $\zs>0.7$ the grain starts
becoming effectively insulating). In Fig. \ref{cflowa}b, 
at $r/R_Q=0.8$ closer to the meeting point (which is at $r/R_Q=0.75$) we see that this strong distinction cannot be made. There is a whole region of parameters to the
right of the black line (marking the theoretically predicted critical flow), in which
the lead-to-lead resistance ($R_{AC}\sim \zeta^2$) decreases, and the
single junction resistance ($R_{AB}\sim \zeta^2+\zs^2$) increases. This
region eventually flows to the Normal fixed point, but in the finite temperature Monte Carlo
simulations this cannot be observed and it appears that an \scs phase exists at 
values $R>R_Q/2$, as shown in Fig. \ref{phasediagram}. 
To the left of the black line in Fig. \ref{cflowa}b we theoretically
expect the \scs phase, but even there we see a crossover which will be
misinterpreted as an FSC phase: $R_{AB}$ decreases and, for a range of
parameters, also $R_{AC}$. $R_{AC}$ only starts to grow at a lower
energy scale, disclosing the true \scs phase. Indeed, in
Fig. \ref{phasediagram} we see that the Monte Carlo calculation
indicates an FSC region at $r/R_Q=0.8$ and $R<0.5R_Q$, where it should be the \scs
phase.

In the ICFP region we again encounter slow crossovers associated with
the intermediate coupling fixed point. As can be seen in
Fig.~\ref{cflowb}c with $r/R_Q=0.7$, again there is a range of parameters which would be
mistaken for the \scs phase, in which $R_{AC}$ seems to drop, while
$R_{AB}$ grows. Since the RG is stopped approximately at the $l_{\text{max}}$ corresponding to our
Monte Carlo calculation, the observed flow in this case is dominated
only by the critical exponent $\lambda_-$ of Eqs.~(\ref{10a}) and (\ref{10b}), which
vanishes at the triple point. For comparison, in Fig.~\ref{cflowb}d we
show the RG flow for $r/R_Q=0.5$. In this case the RG does flow to the
stable NOR and FSC fixed points at energy scales higher than the
lowest temperature accessible in our calculation. Indeed, in this
parameter range there is no longer any evidence of an \scs-like
phase in the Monte Carlo calculation.

\section{The critical two-junction system in the ICFP region - comparison
  with a single Josephson junction}

In this section we investigate the direct NOR-FSC
 transition, and compare the critical behavior of the two-junction system with that
 of a single junction with the same $E_J/E_C$. We make a rather surprising observation: along this phase boundary, the effective resistance of the
 junction, the temperature dependence of the mean phase fluctuations, and the
correlation exponents are within error-bars the same as those in a 
single resistively shunted Josephson junction at criticality (with the same
$E_J/E_C$ and $\Delta\tau E_C$). This suggests that many of the features of the NOR-FSC
transition at the ICFP can be understood in terms of the single
junction Schmid transition, although the ICFP is
an interacting fixed point. We will first present the numerical results and then
 proceed to discuss them in Secs. \ref{RGexp} and \ref{SCexp}.

\subsection{Numerical results and resemblance to the single junction
 \lb{ICFPnums}}

\begin{figure}[t]
\centering
\includegraphics[angle=-90, width=8cm]{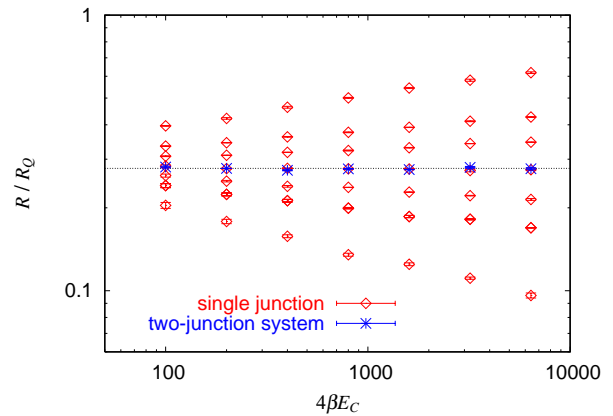}
\caption{Critical lead-to-grain resistance as a function of inverse
temperature for $E_J/E_C=1$ and $\Delta\tau E_C=0.25$. The point
$r=0.5 R_Q$, $R=0.65 R_Q$ on the FSC-NOR phase boundary has been selected (see
Fig.~\ref{phasediagram}). For comparison, we also plot
resistance-versus-temperature data for a single resistively shunted
junction with $E_J/E_C=1$ and $\Delta\tau E_C=0.25$. The values $R_s$
of the shunt resistors are (from top to bottom) $R_Q/R_s=0.9$, 0.95,
0.975, 1.0, 1.025, 1.05, and 1.1, respectively. At the phase
transition point $R_Q/R_s=1$, the resistance of the single junction is
exactly the same as the critical lead-to-grain resistance in the
two-junction system.}
\label{r_two_junctions}
\end{figure}

Associated with the drift of the phase boundary in the ICFP region as a function of $E_J/E_C$ (see Fig.~\ref{boundary_E_J})
is a continuous change in the critical resistance of the junctions and
in the value of the correlation exponents (defined below in Eq.~(\ref{corr})). 
As mentioned above, these features of the fixed point are remarkably
similar in the two-junction model and in a single resistively shunted
junction at criticality.  To illustrate this, we first consider the
critical resistance and plot in Fig.~\ref{r_two_junctions} the
resistance from lead to central grain as a function of inverse
temperature for $E_J/E_C=1$, $r=0.5 R_Q$ and $R=0.65 R_Q$, which is a point on
the FSC-NOR phase boundary. Also shown in the figure is the
temperature dependence of the resistance in a single junction (with
$E_J/E_C=1$ and the same discretization step $\Delta\tau E_C = 0.25$)
for several values of the shunt resistance $R_s$. The curves from top
to bottom correspond to $R_Q/R_s=0.9$, 0.95, 0.975, 1.0, 1.025, 1.05
and 1.1, respectively. For $R_Q/R_s>1$, the junction turns
superconducting as $T\rightarrow 0$, whereas for $R_Q/R_s<1$ it
becomes insulating. At the critical point $R_Q/R_s=1$, the resistance
of the system (junction plus shunt resistor) is precisely the same as
the critical resistance from lead to central grain in the two-junction
model.

\begin{figure}[t]
\centering
\includegraphics[angle=-90, width=8cm]{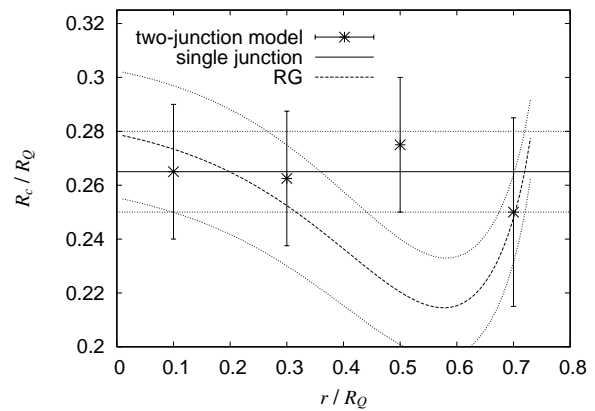}
\caption{Value of the critical lead-to-grain resistance along the
  FSC-NOR phase boundary for $E_J/E_C=1$. The critical resistance is
  nearly independent of $r$ and within error-bars it is the same as in a single junction.
  The solid line shows the critical resistance of the single
  resistively shunted junction with $E_J/E_C=1$ and $R_s =
  R_Q$. Error estimates are indicated by the dotted
  lines.  The dashed line is the result of an RG-based calculation of the effective
  lead-to-grain resistance. 
 }
\label{r_critical}
\end{figure}

By and large, the critical resistance
of the two-junction system does not vary along the FSC-NOR phase boundary for
fixed Josephson coupling. We plot its value in Fig.~\ref{r_critical}
as a function of $r$ and compare it with the corresponding result for
a single resistively shunted junction. A very good agreement is
evident for $r/R_Q=0.1,\,0.2$, and $0.5$. For $r/R_Q=0.7$, closer to the
meeting point of the three phases, the agreement is less good although
still within error bars.

\begin{figure}[t]
\centering
\includegraphics[angle=-90, width=8cm]{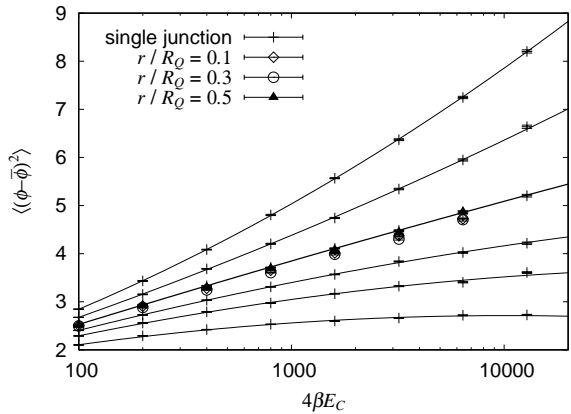}
\caption{Phase fluctuations $\langle(\phi-\bar\phi)^2 \rangle$ as a
  function of inverse temperature. The curves from top
  to bottom show the data for a single resistively shunted junction
  with $R_Q/R_s=0.9$, 0.95, 1.0, 1.05, 1.1, and 1.2 respectively. The
  symbols show the same quantity calculated in the two-junction model
  at criticality. Three points along the FSC-NOR phase boundary have
  been chosen and all the data agree with those of the single junction
  with identical values of $E_J/E_C=1$ and $\Delta\tau E_C=0.25$. }
\label{phase_fluctuations}
\end{figure}

Another quantity we measured is the mean phase fluctuation
$\langle (\phi-\bar\phi)^2 \rangle$, which we found in
Ref.~\onlinecite{junction} to grow proportional to the logarithm of
the inverse temperature at criticality. The same is
true in the two-junction system. In Fig.~\ref{phase_fluctuations} we 
plot $\langle(\phi-\bar\phi)^2 \rangle$ as a function of inverse temperature. The lines show the data for a single junction
with the bold line marking the logarithmic growth of the phase
fluctuations at the critical point. The diamonds, circles and triangles show
the data obtained from the two-junction system at criticality. They
correspond to points along the FSC-NOR phase boundary with $E_J/E_C=1$ and $\Delta\tau E_C=0.25$,
as in the single junction case. The uncertainty on the critical value
of $R$ has not been taken into account in the error bars of
Fig.~\ref{phase_fluctuations}. This uncertainty can account for the small
deviations between the different data sets.

\begin{figure}[t]
\centering
\includegraphics[angle=-90, width=8cm]{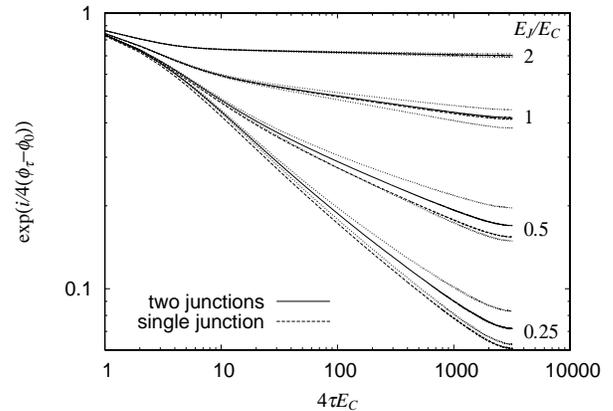}
\caption{Correlation function (\ref{corr}) for $q=0.25$ and $\beta
  E_C=1600$. The curves correspond to $r=0.5R_Q$, and from top to bottom:
  $E_J/E_C=2$, $1$, $0.5$ and $0.25$, respectively; $R$ in these plots
  is tuned to the FSC-NOR critical line. The solid lines
  show the exponents computed for the two-junction system and the
  dashed lines those obtained for a single resistively shunted
  Josephson junction at criticality. The dotted lines indicate the
  error on the two-junction calculation originating from the
  uncertainty on the critical resistances.}
\label{eta}
\end{figure}

In addition, we considered the exponents $\eta(q)$, which we measured for
several values of $E_J/E_C$ at $r=0.5R_Q$ and with $R/R_Q$
corresponding to the FSC-NOR phase boundary.  
We define the latter as in Ref.~\onlinecite{Tewari} using the correlation function
\begin{equation}
\text{corr}_q(\tau) \equiv \langle \exp[iq(\phi(\tau)-\phi(0))] \rangle,
\label{corr}
\end{equation}
with $q$ some non-integral real number and $\phi$ the phase difference
across one of the junctions. In the normal phase and at criticality the
correlations decay as
\begin{equation}
\text{corr}_q(\tau) \sim \tau^{-2\eta(q)}. 
\end{equation}
In the superconducting phase one observes a power-law decay of the
connected correlation function
\begin{equation}
\text{corr}_q(\tau) \equiv \langle \exp[iq(\phi(\tau)-\phi(0))] \rangle-|\langle \exp[iq\phi(\tau)]\rangle|^2.
\label{corr_conn}
\end{equation}
In Fig.~\ref{eta} we plot $\text{corr}_{q=1/4}(\tau)$  for the
Josephson coupling strengths $E_J/E_C=0.25$, 0.5, 1, 2 (as in
Fig.~\ref{boundary_E_J}) and the corresponding critical resistances
$r=0.5$ and $R=0.5375(200)$, 0.5875(150), 0.650(20), 0.6875(200),
respectively. We compare these correlation functions to those obtained
for a single junction at criticality ($R_\text{shunt}=R_Q$) for the
same values of $E_J/E_C$ and the same discretization step $\Delta\tau
E_C=0.25$. As can be seen in Fig.~\ref{eta}, the correlation functions,
and thus also the critical exponents $\eta(q)$ perfectly agree for
$E_J/E_C=2$ and 1, 
while they agree within error bars for $E_J/E_C=0.5$ (the dotted lines
in the figure show the correlation functions computed at
$R=R_\text{critical}\pm \text{error}$). For
$E_J/E_C=0.25$ the agreement is no longer as good, but the exponents for
a single junction and for two junctions remain close and well within error-bars. 

While the exponent $\eta(q)$ varies as a function of $E_J/E_C$ in
a remarkably similar way as in a single junction, there is no dependence
of this exponent for fixed $E_J/E_C$ on the values of $r$ and $R$,
that is on the position along the FSC-NOR phase boundary shown in
Fig.~\ref{phasediagram}. Remember that also the
effective lead-to-grain resistance was within error-bars independent of $r$ on the
FSC-NOR phase boundary (Fig. \ref{r_critical}). 

\begin{figure}[t]
\centering
\includegraphics[angle=-90, width=8cm]{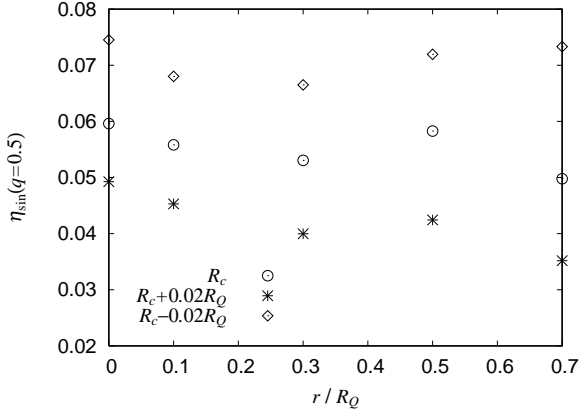}
\caption{
Correlation exponents $\eta_{\sin}(q=0.5)$ obtained from a fit to
$f(\tau)=a/\sin(\pi/\beta\tau)^{2\eta_\text{sin}}$ along the FSC-NOR phase boundary (see
Fig.~\ref{phasediagram}) for constant $E_J/E_C=1$ and $\beta
E_C=1600$. The main source of error is the uncertainty of $\pm
0.02R_Q$ on the critical resistance $R_c$ for a given value of
$r$. The independence of the exponent $\eta_{\sin}(q=0.5)$ with
respect to $r$ and the constant 
lead-to-grain resistance shown in Fig.~\ref{r_critical} indicate that the critical properties only depend on the value of $E_J/E_C$.
}
\label{fit_EJ}
\end{figure}

We plot $\eta_{\sin}(q)$, obtained from a fit to
\begin{equation}
f(\tau)=\frac{a}{\sin((\pi/\beta)\tau)^{2\eta_{\sin}}},\label{fit_sin}
\end{equation}
along the FSC-NOR phase boundary for $q=0.5$, $\beta E_C=1600$ and
$E_J/E_C=1$ in Fig.~\ref{fit_EJ}. The measured exponents range between
0.05 and 0.06 and are thus well within error-bars. To estimate the
error, which mainly originates from the uncertainty of $\pm 0.02R_Q$
on the critical value of $R_c$ (for a given $r$), we also plot the
exponents measured at $R_c\pm 0.02R_Q$ in the figure.

\subsection{RG analysis of the critical effective resistance in the
  two-junction system \lb{RGexp}}

The extensive investigation in the previous Sec.~\ref{ICFPnums}
demonstrated a remarkable resemblance between the behavior at
criticality of the two-junction system and a single junction with the
same $E_J/E_C$. In this section we will account for this resemblance
by comparing the measured effective resistance at criticality of the two-junction
system with the resistance as predicted from the RG flow equations (\ref{s2a}) and (\ref{s2b}). We emphasize that such an analysis is only approximate,
since Eqs.~(\ref{s2a}) and (\ref{s2b}) are derived in the limit of large $E_J/E_C$, and
the case we are concentrating on is that of an intermediate
$E_J/E_C$. Nevertheless, the strong-coupling RG equations provide a
rather good fit to the behavior of the system at criticality.
Since all quantities considered in Sec. \ref{ICFPnums} (lead-to-grain
resistance, phase fluctuations, and phase correlations) are related to
the effective lead-to-grain resistance, the analysis of the latter
suffices.

The first step in the analysis is to connect the measured
lead-to-grain effective resistance with the resistances $R$ and $r$,
and with the phase-slip fugacities $\zeta$ and $\zs$ (which we will
later substitute with their fixed-point values, $\zeta^*$ and $\zs^*$). This amounts to
solving the circuit shown in Fig. \ref{ICFPcircuit}.  The junctions in
the figure were each replaced by two components: a resistor 
\be
R{_\zeta}\approx \alpha \zeta^2,
\lb{r1}
\ee
and a voltage drop due to phase-slip
dipoles, $V_D$. The latter is given by 
\be
V_D\approx \alpha \zs^2
(I_1-I_2).
\lb{r2}
\ee
The solution of the circuit in Fig.~\ref{ICFPcircuit} is sketched in
App.~\ref{appC}. It leads to the following
expression for the lead-to-grain resistance
\be
R_\text{eff}=\frac{(R^2+2rR)(\zeta^2 + \zs^2)
  +(R+r)\alpha\zeta^2(\zeta^2 + 2\zs^2)}{(R/\alpha + \zeta^2)(2r + R +\alpha\zeta^2 + 2\alpha\zs^2)}.
\lb{r3}
\ee

\begin{figure}
\includegraphics[width=7.5cm]{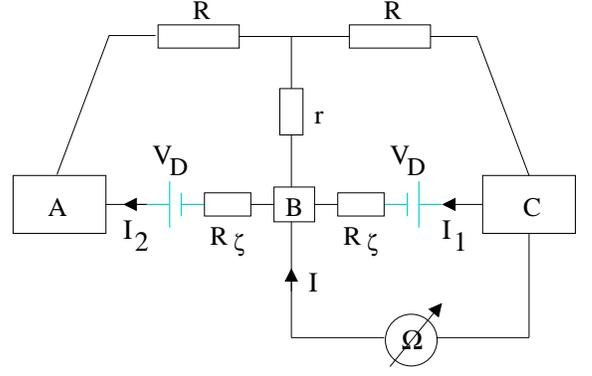}
\caption{A circuit describing the measurement of the lead-to-grain
  resistance. Each junction is replaced by two components: a resistor
  $R_{\zeta}=\alpha\zeta^2$ which describes the response of single
  phase-slips, and a voltage drop $V_D$ which in junction BC is
  $\alpha\zs^2(I_1-I_2)$ and in junction AB is
  $\alpha\zs^2(I_2-I_1)$. The Ohm-meter is considered as a current
  source, $I$.  \lb{ICFPcircuit}}
\end{figure}
 
In Eqs. (\ref{r1}) and (\ref{r2}) we used the coefficient $\alpha$ to
convert the square of the phase-slip fugacities to resistances. This
coefficient is going to be the only input into the RG flow analysis of
the effective lead-to-grain resistance. An estimate for this coefficient
can be obtained from the data point at $E_J/E_C=1$ and $r/R_Q=0.5$ in
Fig. \ref{r_critical}, in conjunction with the measured single-junction
effective resistance. The critical shunt resistance at this point is
$R/R_Q=0.65 \pm 0.02$. From the integrated RG flow we can find $\zeta_0$ such
that for $r=0.5R_Q$ and $R=0.65R_Q$ the point $(\zeta_0,\,0)$ lies on
the critical manifold. We find
\begin{equation}
\zeta_0=0.295\pm 0.01.
\end{equation} 
This $\zeta_0$ corresponds to $E_J/E_C=1$. 
Note that $\zeta_0$ is a function of $E_J/E_C$ only in the strong
coupling limit, and should only depend weakly on the resistances
$r,\,R$ (c.f. Ref. \onlinecite{Schoen}). The same applies to
the coefficient $\alpha$. Bearing this in mind, we can calculate
$\alpha$ by using the measured resistance of the single-junction
system. 
This measured resistance is the resistance of the junction, $\alpha\zeta^2$, parallel to the shunt resistor, which equals $R_Q$ at criticality.
Thus, for $E_J/E_C=1$ we have (Fig.~\ref{r_critical})
\begin{equation}
\frac{\alpha\zeta_0^2\cdot R_Q}{R_Q+\alpha\zeta_0^2}=0.265\pm0.015
\label{r3.5}
\end{equation}
from which we find
\be
\alpha=4.15\pm 0.35.
\lb{r4}
\ee

Now we have all the pieces to predict the effective lead-to-grain
resistance, $R_\text{eff}$ on the FSC-NOR critical line $E_J/E_C=1$. By using
Eq.~(\ref{r3}) with the fixed-point values $\zeta^*$ and $\zs^*$
from Eqs.~(\ref{i18}), and $\alpha$ from Eq.~(\ref{r4}), we obtain the
curve of $R_\text{eff}$ as a function of $r$. 
The result of
this calculation is shown in Fig.~\ref{r_critical} by the dashed line. One can see that the resistance predicted by the RG changes little in the entire range, and remains in reasonably
good agreement with the observed Monte Carlo $R_\text{eff}$. 

Note that we
assumed that the measured resistance is due to the fixed point
characteristics of the two-junction system. From Sec. \ref{COsec}, however,
we know that in the vicinity of $r=0.75R_Q$, where the three phases
meet, crossover effects are dominant. Therefore we expect the
lead-to-grain resistance calculated in this section to deviate from the measured $R_\text{eff}$ in that
vicinity. Another caveat for the current calculation is that it is
correct up to second order in the phase-slip
fugacities; fourth-order contributions to $R_\text{eff}$ are neglected
(although we keep fourth order terms in Eq.~(\ref{r3})). These
corrections may also account for deviations from the measured
$R_\text{eff}$. 


\section{The ICFP as a self-consistent fixed point \lb{SCexp}}

\begin{figure}[t]
\includegraphics[angle=0,width=8.5cm]{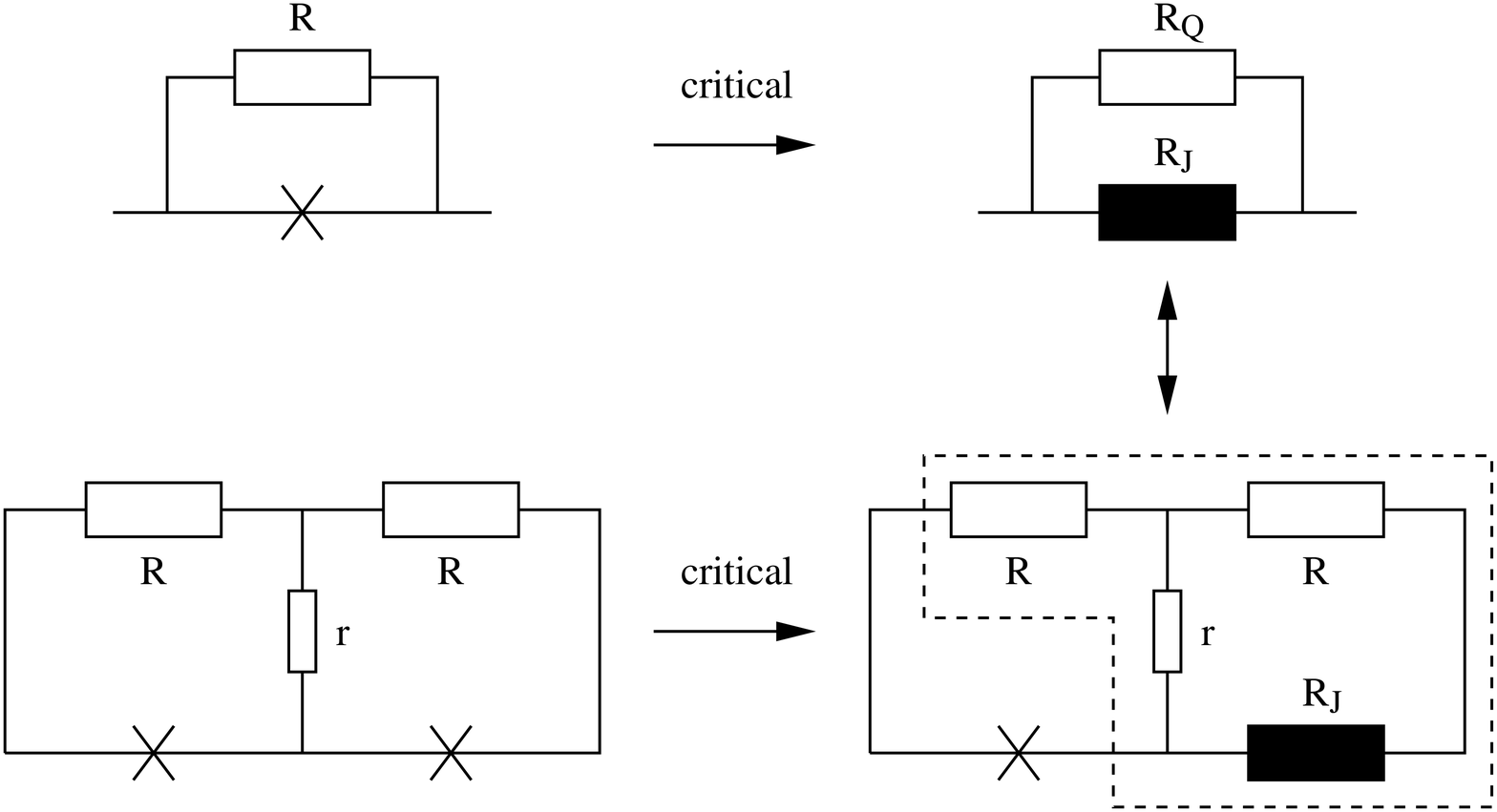}
\caption{Illustration of the mean-field theory for the FSC-NOR
  transition in the two-junction model. A Josephson junction at
  criticality is replaced by a resistor $R_J$, such that the
  resistance of the circuit on the top right corresponds to the
  measured resistance of the device. Replacing one of the junctions in
  the two-junction system by $R_J$, one can calculate the effective
  shunt resistance $R_s^\text{eff}$ of the other junction. Identifying
  $R_s^\text{eff}$ with $R_Q$ yields an equation for the FSC-NOR phase
  boundary.}
\label{fig_mean_field}
\end{figure}

In the previous section we investigated the FSC-NOR transition
extensively, and demonstrated a remarkable resemblance between the 
two-junction and single-junction systems at criticality. We adequately
explained this surprising resemblance using the RG from Sec.~II. In this
section, however, we provide yet another explanation, albeit ad-hoc,
for this resemblance.  The alternative explanation is that the ICFP
can be approximated as a self-consistent fixed point. In such a
mean-field theory, illustrated in Fig.~\ref{fig_mean_field}, the
physics of a single junction emerges naturally. 
The idea behind this approach is that a Josephson junction undergoes a SC-NOR transition when its 
effective shunting resistance is $R_Q$. 

\begin{figure}[t]
\includegraphics[angle=-90,width=8cm]{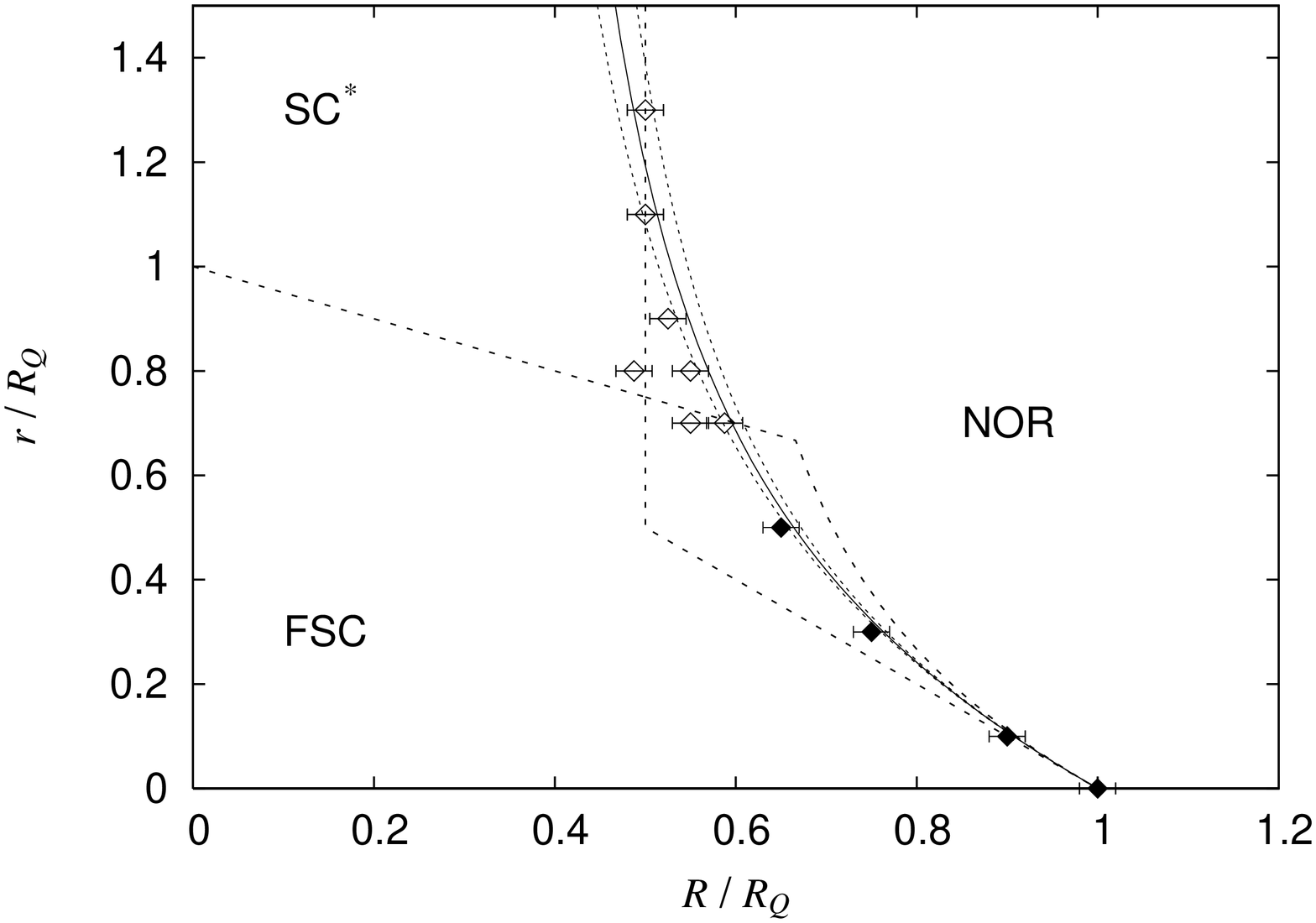}\\
\includegraphics[angle=-90,width=8cm]{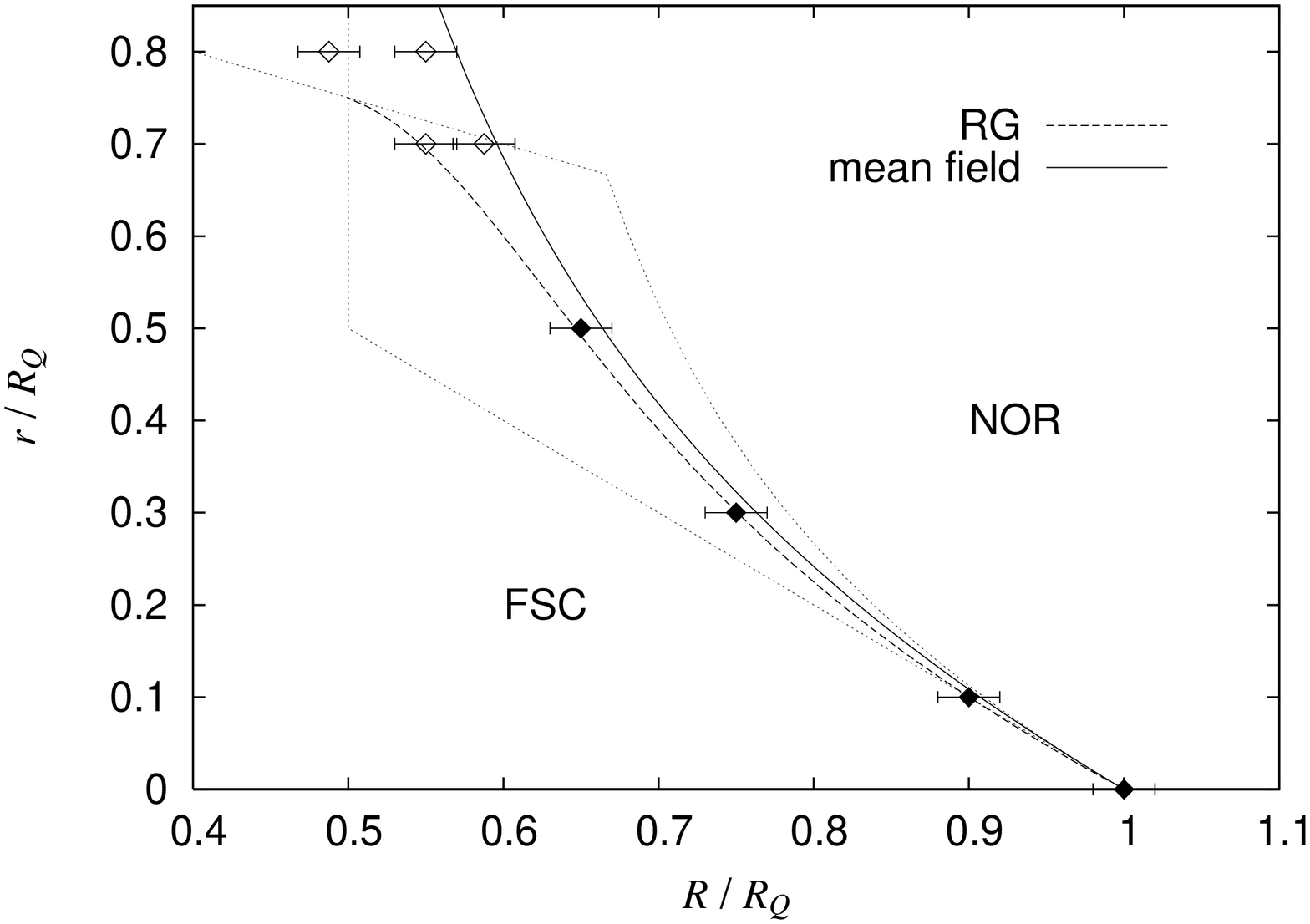}
\caption{Upper figure: SC-NOR phase boundary obtained from the mean-field prediction Eq.~(\ref{mean_field}) for $E_J/E_C=1$. The critical resistance $R_c$ of the single junction with $R_s=R_Q$ is $R_c/R_Q\approx0.265$ (see Fig.~\ref{r_critical}). The dotted lines show how the error on $R_c$ influences the outcome. A remarkably good agreement with the Monte Carlo results is obtained. \\
Lower figure: A closeup of the phase diagram in the ICFP region for $E_J/E_C=1$ ($\zeta_0=0.295$). The solid line is the mean-field phase boundary from Eq.~(\ref{mean_field}) and the dashed line is the RG prediction for the FSC-NOR phase boundary. The RG-results agrees well with the mean-field prediction and with the Monte Carlo results for $r<0.6R_Q$. Filled dots show the position of the FSC-NOR transition, open dots indicate a SC-SC$^*$ or SC$^*$-FSC transition, respectively. See section IVB for a discussion of crossover effects.}
\label{MFTRGpd}
\end{figure}

\begin{figure}
\includegraphics[width=7.5cm]{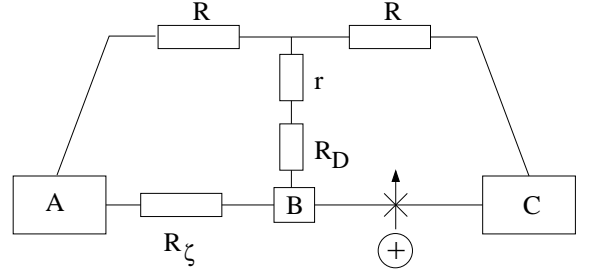}
\caption{Effective circuit in the mean-field approximation - taking
  phase-slip dipoles into account. Junction
  AB is approximated by a resistor with resistance
  $R_{\zeta}=\alpha\zeta^2$ which is due to single phase slips. In
  addition, phase-slip dipoles add $R_D=\alpha\zs^2$ to the resistor $r$. The
  location of $R_D$ is determined such that it responds to the
  difference in the current on the two junctions
  (c.f. Eq.~(\ref{r2})). This circuit is equivalent to the one shown in
  Fig. \ref{ICFPcircuit} with $I=0$. In the mean-field approximation
  we assume that a NOR-FSC phase transition occurs when the effective
  shunting resistance that a phase slip on the junction BC probes
  equals $R_Q$. Furthermore, we assume that $R_D$ is negligible - an
  assumption which is well justified with respect to the Monte Carlo
  calculation in the ICFP region and near the
  meeting point of the three phases, where crossover effects are
  evident (c.f. Fig. \ref{fig_mean_field}). \label{mft}}
\end{figure}

Consider the two-junction system with phase slip fugacities $\zeta_0$ (with $\zeta_D=0$) in both junctions. 
In mean-field, one junction sees the other as an effective resistor (Figs. \ref{fig_mean_field} and \ref{mft}) with resistance 
\begin{equation}
R_J=\alpha\zeta_0^2,
\end{equation}
where $R_J$ is defined from Eq.~(\ref{r3.5}). 
Therefore the effective shunting resistance on each of the junctions is 
\be
R_s\approx R+\frac{r\l(R+R_J\r)}{r+R+R_J}
\lb{r5}
\ee
and criticality is obtained when 
\be
R_s=R_Q.
\lb{r6}
\ee
From Eqs.~(\ref{r5}) and (\ref{r6}) we find an expression for the FSC-NOR phase boundary,
\be
r=\frac{(R+R_J)(R_Q-R)}{2R+R_J-R_Q}
\lb{mean_field}
\ee
as a function of the parameter $R_J$, which depends on $E_J/E_C$. The limiting cases for strong and weak Josephson potential work out correctly. For $E_J/E_C\to \infty$, $R_J\to 0$ (superconducting junction) and Eq.~(\ref{mean_field}) reduces to
\be
r = \frac{R(R_Q-R)}{2R-R_Q},
\ee  
whereas for $E_J/E_C\to 0$, $R_J\to \infty$ (insulating junction) and we obtain
\be
r = R_Q-R.
\ee
These are indeed the equations describing the FSC-NOR phase boundaries shown in Fig.~\ref{phasediagram_refael}.

In the mean-field approximation, each junction at criticality behaves as
an {\it independent} junction, and thus exhibits the same phase correlations
and fluctuations as the single junction with the same $R_J$, and
hence exhibits the same dependence on $E_J/E_C$. 

Although this explanation seems naive, it does allow a simple and quantitative
understanding of the observed (and calculated) properties of the two
junction-system at the SC-NOR transition point. As further evidence,  in
Fig.~\ref{MFTRGpd}, we compare the phase boundaries in the ICFP region
as obtained  by the simple mean-field treatment
(Eq.~(\ref{mean_field})), the RG calculation (explained in App.~\ref{appB}), and the Monte Carlo results.
For $r/R_Q<0.7$ there is a good agreement between the three approaches. In the vicinity of the
meeting point of the three phases, the RG boundary traces the measured FSC-\scs phase boundary, while the mean field prediction agrees with the Monte Carlo
\scs-NOR phase boundary, even for values of $r>0.75R_Q$.
Note, though, that despite the agreement with the Monte Carlo results, the mean-field argument does not apply to the \scs-NOR phase boundary. 

The  good agreement of the mean-field boundary with the measured
FSC-NOR boundary seems rather surprising. It can be explained once we understand why the
mean-field treatment is (at least qualitatively) reasonable.  The
mean-field approach makes two assumptions: (1) a junction with
phase-slip fugacity $\zeta$ can be replaced by a resistor
$\alpha\zeta^2$, and  (2) $\alpha\zs^2$ can be neglected compared to
$r$ (see Fig. \ref{mft}).  At this point we can only justify the first
assumption on qualitative, but not on analytical, grounds. Assumption
(2) on the other hand is justified by the fact that the fixed point
values of the  phase-slip fugacities obey $\alpha(\zs^*)^2\ll r$ for
much of the ICFP region. For $\zeta_0=0.295$ (corresponding to
$E_J/E_C=1$)  this ratio rises slowly from zero at $r=0$ to $0.25$ at
$r=0.6R_Q$; it crosses $1$ at   $r=0.73R_Q$. In the region where
$\alpha(\zs^*)^2$ is no longer negligible,  i.e., $r>0.6R_Q$,
crossover effects set in, and the RG flow is cut off at the lowest
energy scale of the Monte Carlo simulation. In this case $\zs$ does not
reach its  fixed point value and will be observed as small. The
crossover effects justify assumption (2) of the mean-field in the same
way that they explain the deviations from the  predicted phase diagram
in Sec. \ref{COsec}. Thus the mean-field approach reproduces 
the phase diagram with cross-over effects. 

Above we made the assumption that in the mean-field approach  a
junction undergoes the SC-NOR transition when the effective shunting
resistance sees equals $R_Q$. We also derived the phase boundary
using the RG flow (App.~\ref{appB}) without making any such
assumption. This raises the question: how close to $R_Q$ is the effective shunting
resistance along the RG-predicted phase boundary? This effective
shunting resistance is given by
\be
R_s^\text{eff}\approx
R+\frac{\l(R+R_{\zeta}\r)\l(r+R_D\r)}{r+R+R_{\zeta}+R_D},
\lb{r7}
\ee
which can be deduced by observation from Fig.~\ref{mft}, with
$R_D=\alpha\zs^2$ and $R_{\zeta}=\alpha\zeta^2$. 
In Fig.~\ref{effectiveshunt} we plot this effective shunting
resistance for $\zeta_0=0.295$, which corresponds to $E_J/E_C=1$, as a
function of $r$ for the entire ICFP region, $0<r/R_Q<0.75$. We assume that
$\zeta$ and $\zs$ take their fixed point values from
Eqs.~(\ref{i18}). As can be seen, the effective shunting resistance stays
very close to $R_Q$ over the entire range. At this point we are
unable to say whether this is just a coincidence (as the success of
the mean-field treatment above may therefore be), or a universal
property of the phase boundary, which the approximate RG roughly
reproduces. To answer this question, a more thorough consideration of
the sine-Gordon model is necessary. 

\begin{figure}
\includegraphics[width=7.5cm]{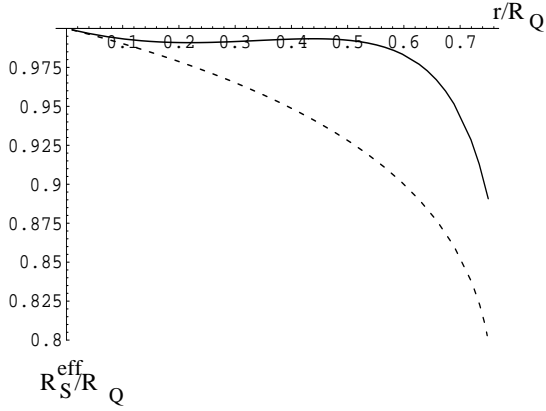}
\caption{The RG-based calculation of $R_s^\text{eff}$ (solid line) given in Eq.~(\ref{r7}) as a function of
  $r/R_Q$ on the phase boundary in the ICFP region with $E_J/E_C=1$
  ($\zeta_0=0.295$). The effective shunt resistance is very close to $R_Q$
  in the entire region, giving more validity to the proposed
  mean-field approach to the double sine-Gordon model of the
  two-junction system. The dotted line is the effective shunt
  resistance when we set $R_{\zeta}=0$ and $R_D=0$, and is given for reference (this is just
  $R+rR/(r+R)$). \lb{effectiveshunt}}
\end{figure}


\section{Correlation function in the NOR phase \lb{CNOR}}

\begin{figure}[t]
\centering
\includegraphics[angle=-90, width=8cm]{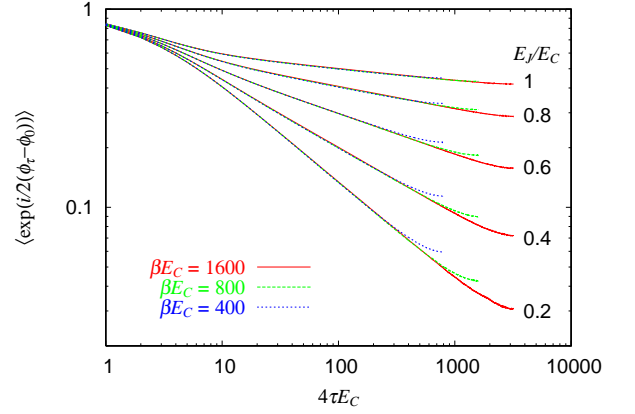}
\caption{Correlation functions $\text{corr}_{q=0.5}(\tau)$ in the FSC phase for different temperatures and values of $E_J/E_C$, but fixed resistance $r=0.5 R_Q$ and $R=0.65 R_Q$ (corresponding to $(E_J/E_C)_\text{critical}=1$). Different $E_J/E_C$ yield considerably different exponents $\eta(q)$.}
\label{r_E_J}
\end{figure}

\begin{figure}[t]
\centering
\includegraphics[angle=-90, width=8cm]{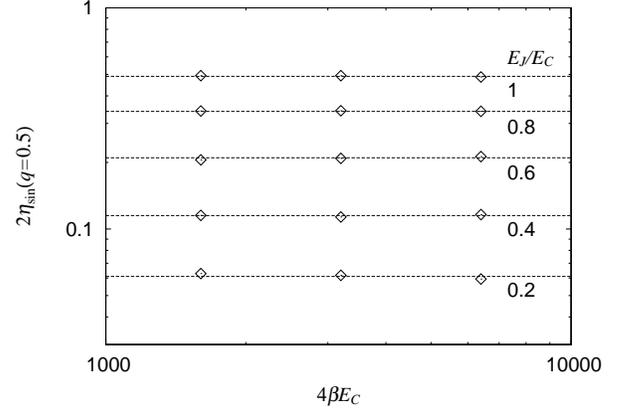}
\caption{Correlation exponents $\eta_{\sin}(q=0.5)$ obtained from a fit to Eq.~(\ref{fit_sin}) for the temperatures $\beta E_C = 1600$, 800 and 400. }
\label{eta_beta}
\end{figure}

In this section we compute and briefly consider the decay of the
correlation function (see Eq.~(\ref{corr}))
\[
\text{corr}_q(\tau) \equiv \langle \exp[iq(\phi(\tau)-\phi(0))] \rangle
\]
in the NOR phase of the system. 
We set $r=0.5 R_Q$ and $R=0.65 R_Q$, and calculate $\text{corr}_q(\tau)$ for different
values of the Josephson coupling energy. This choice of $r$ and $R$ corresponds to a
critical value $(E_J/E_C)_\text{critical}=1$ (see
Fig.~\ref{boundary_E_J}). Our results are shown in
Fig.~\ref{r_E_J}. Interestingly, each curve has an approximate
power-law behavior, but with exponents that vary with $E_J/E_C$. 
The correlation exponents $\eta_{\sin}(q=0.5)$ obtained from a fit to Eq.~(\ref{fit_sin}) are shown in Fig.~\ref{eta_beta} for the temperatures $\beta E_C = 1600$, 800 and 400. This behavior is very different from that seen in a single resistively shunted junction, where different $E_J/E_C$ (at fixed shunt resistance $R_s$) yield the same exponents $\eta$ in the NOR phase\cite{junction}.

However, one point should be kept in mind:
even though in the previous section, we demonstrated and discussed the similarities
between the two-junction model and the single junction at
criticality, there are also stark differences between the two systems. 
While in the single junction, flow lines
corresponding to systems with the same shunt resistance but different
$E_J/E_C$ merge (as they flow parallel to the $E_J/E_C$ axis only), in the
two-junction system, under the same circumstances, they do not overlap (see
Fig \ref{fig4}). Indeed, in  Eqs.~(\ref{w20b}) and
(\ref{s2b}) we need an extra parameter, $J_+$ and $\zs$, in order to
write the renormalization group flow equations for the two-junction
system.

In the {\it single-junction} model we expect systems with different
$J\propto E_J/E_C$ but same shunt resistance to show the same low energy
behavior. More specifically, at weak $J$ we expect
\be
\text{corr}_q\l(\tau,\,J\r)=b^{-2q^2 \frac{R}{R_Q}}\text{corr}_q\l(\frac{\tau}{b},\,\frac{J}{b^{R/R_Q-1}}\r).
\lb{cr1}
\ee
To compare two different systems, with junctions $J_1$ and $J_2$, we
can choose
\be
b=\l(\frac{J_1}{J_2}\r)^{1/\l(R/R_Q-1\r)}
\ee
and obtain
\begin{align}
&\text{corr}_q\l(\tau,\,J_1\r)=\nonumber\\
&\hspace{5mm}\l(\frac{J_1}{J_2}\r)^{-2q^2 R/(R-R_Q)}\text{corr}_q\l(\tau\Big(\frac{J_1}{J_2}\Big)^{1/\l(R/R_Q-1\r)},\,J_2\r).
\lb{cr2}
\end{align}
If the decay of the correlations is a power law,
$\text{corr}_q\l(\tau\r)\sim |\tau|^{-2\eta(q)}$, then the scaling
relation (\ref{cr2}) clearly shows that $J$ must drop out of the power
law. In fact, from Eq. (\ref{cr1}) we find directly
\be
\eta(q)=q^2 \frac{R}{R_Q}.
\lb{cr3}
\ee

In the two-junction system, we observe a power law decay of
$\text{corr}_q\l(\tau,\,J_1\r)$, but with $\eta(q)$ which seemingly depends also
on $J$. 
In the absence of our understanding of the RG flows of the
two-junction system, we might guess that a scaling relation similar to
Eq. (\ref{cr1}) holds, but with the shunt resistors, $R$ and $r$ also
flowing. This possibility, however, seems to contradict Ref. \onlinecite{Bulgadaev}, which claims that in sine-Gordon models the resistance can not get
renormalized due to its singular $|\omega|$-frequency dependence.

We can, however, qualitatively explain the dependence
of $\eta(q)$ on $J$ in terms of the RG flow equations (\ref{eq14b}). 
In terms of the two-junction RG flows for $J$ and $J_+$ (or $\zeta$
and $\zs$), a scaling form such as Eq.~(\ref{cr2}) does not hold in
the two-junction system in the ICFP region. Instead, we need to
integrate Eqs.~(\ref{eq14a}) and (\ref{eq14b}). In Fig.~\ref{NORflows}, the RG
flows corresponding to $R=0.65R_Q$, $r=0.5R_Q$, and
$J_0=0.5,\,0.4,\,0.3,\,0.2$ ($E_J/E_C=1,\,0.8,\,0.6,\,0.4$), are
shown. In the accessible energy range, all flows take
different paths in the $(J,\,J_+)$ space. Also, they are all far from
saturating at $J=J_+=0$. The latter observation provides the reason
for the slower decay of the correlations for higher $J$: in the
accessible energy range the two-junction system seems to have a larger
effective $J$ and $J_+$, the higher the bare $J_0$ is. These finite $J$ and
$J_+$, in turn, provide extra conductance in the system, and thus
strengthen phase coherence, and suppress the decay of correlations. 

We furthermore note that the flow in the temperature range $400<\beta E_C < 1600$ (corresponding to $6<l<7.4$) is rather slow, which might explain why there is hardly any temperature dependence in the measured $\eta$ (Fig.~\ref{eta_beta}). However, we cannot rule out a scenario where the flows have actually converged to a set of distinct fixed points (with e.g. different values of $J_+$), although the origin of such additional fixed points remains unclear. 

\begin{figure}
\includegraphics[width=7.5cm]{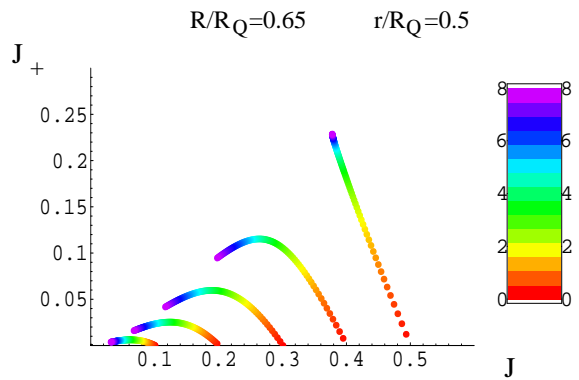}
\caption{The integrated weak-coupling RG flows of Eqs. (\ref{eq14b})
  in the $(J,\,J_+)$ plane for $R=0.65 R_Q$ and $r=0.5 R_Q$. The value
  $J_0=0.5$ corresponds to the critical value $E_J/E_C=1$. These flows
  show qualitatively how the paths with different bare coupling
  $J_0$ essentially do not overlap, and also do not reach zero in the
  accessible energy range. The flows are color coded to show the
  evolution of the logarithmic RG scale $l=\ln \Lambda_0/\Lambda$
  along the flow ($\Lambda_0$ is the bare UV cutoff and $\Lambda$ is
  the renormalized cutoff). In Fig. \ref{r_E_J} $l\leq 8$, as in this
  plot. \lb{NORflows}}
\end{figure}

\section{Conclusions}

In this paper, we presented an RG analysis and simulation results for a
symmetric two-junction system. The phase diagram was found to contain
all three phases expected by theory: (i) fully superconducting (FSC), (ii) normal (NOR) and
(iii) superconducting from lead to lead but normal from lead to grain
(\scs). Furthermore, the phase boundaries obtained by the Monte Carlo simulation agreed
with the theoretical prediction, except for deviations near the
tricritical point
where the three phases meet. These deviations were interpreted as finite
temperature crossover effects, and give indirect evidence for 
the existence of the intermediate coupling fixed point.
The behavior of correlation functions in the NOR phase may also be indirect evidence for the flow of the coupling $J_+$, as predicted by the RG-theory.

The two-junction system possesses the unusual ICFP region, in which
the NOR-FSC transition can be tuned by $E_J/E_C$ as well as by the
dissipation. We calculated this phase boundary using the RG
treatment of Ref. \onlinecite{Refael1}, and found that it fits the Monte Carlo result quite well
(c.f. Fig. \ref{close-up}). Theoretically it was shown that in this
region the properties of the two-junction system are determined by a
new intermediate coupling fixed point, with continuously varying
exponents. 

An efficient Monte Carlo algorithm was used to thoroughly
investigate the
intriguing ICFP region and the NOR-FSC phase transition. We calculated
the lead-to-grain resistance, average phase fluctuations, and
phase correlations for several locations on the
critical NOR-FSC manifold. This manifold is two dimensional, and can
be parameterized by $r$ and $E_J/E_C$. 

For the various sets of $(r,\,E_J/E_C)$ on the NOR-FSC critical
manifold we made a surprising observation:  the effective fixed point
lead-to-grain resistance, $R_{AB}$, as
  well as other critical properties of the two-junction model, were
  nearly identical to those of a single junction (with
the same $E_J/E_C$) at criticality ($R_s=R_Q$). Moreover, the
effective lead-to-grain resistance is within error-bars independent of $r$ and $R$, and depends only on $E_J/E_C$. 

Given the success of the RG analysis of the two-junction model, we
tried to use it in order to account for the above observations. Using simple scaling arguments, we extended the RG of
Ref. \onlinecite{Refael1} and Sec. II to compute the effective
lead-to-grain resistance at the critical point. This calculation also
yielded a nearly $r$ independent effective resistance, and since in
the limit $r=0,\,R=R_Q$, the two-junction system and the single
junction system coincide, the observations about the resemblance of
the two systems seem to be satisfactorily explained. The approximate RG-based
resistance calculation fitted the QMC results quite well (Fig. \ref{r_critical}). 

But the observed similarity of the single- and two-junction
systems could also be qualitatively interpreted in a more interesting
way. The observation can be taken to indicate that at criticality, each of the two junctions in
the system sees an effective environment that imitates an $R_s=R_Q$
shunt, and therefore shows the same effective fixed point resistance
and critical properties
as a single isolated junction at criticality. This mean-field
argument was developed in Sec. \ref{SCexp}, 
and it allowed us to predict the location of the FSC-NOR phase boundary at
intermediate Josephson energy well within error bars
(Fig. \ref{MFTRGpd}). In addition, we used the RG predictions to compute the
effective shunt resistance for each junction, and we found an
astonishing result: In most of the ICFP, the effective shunt for
each junction does not deviate more than 1\%  from $R_Q$ (Fig.\ref{effectiveshunt}).

The apparent success of a mean-field theory in this interacting system
seems quite remarkable. Even more so is the fact that the RG-based
calculation of the effective shunt on the critical manifold also
yields with good accuracy $R_Q$. This is indirect evidence for an
internal structure in the coupled sine-Gordon theory that describes
the two-junction system. If this approximation turns out to be
generally valid, it may also give insight into the solution of
coupled sine-Gordon models at intermediate couplings where no expansions can
be done and little is analytically known. 

The relevance of our work goes well beyond the two junction
system. A closely related class of systems consists of 
resistively shunted Josephson junction arrays of two or more
dimensions (see e.g. Ref.~\onlinecite{Kivelson-Chakravarty}).
The RG equations derived for the two-junction system \cite{Refael1} were
shortly thereafter also derived for a two-dimensional triangular array of
resistively shunted Josephson junctions in the weak coupling regime:
in Ref. \onlinecite{Tewari}, Tewari {\it et al.} show that the weakly
coupled triangular array undergoes a superconductor-to-metal transition
almost identical to the direct NOR-FSC transition of the two-junction
system (the RG equations for the two systems differ by one coefficient). As you may recall, this transition is in the interesting and
novel ICFP region. Tewari {\it et al.} also considered the square lattice,
in which the RG equation are third-order in the Josephson
coupling, a difference which is unimportant near the ICFP. Hence, the work presented in this paper also verifies many of the results in
Ref. \onlinecite{Tewari}. 

In the future, we plan to adapt the Monte Carlo algorithm such that it can be applied 
to sizable arrays of Josephson junctions. This would allow
for the first time direct controlled investigation of the
Kosterlitz-Thouless NOR-\scs (normal-superconducting) transition  in
one-dimensional resistively shunted and free Josephson-junction
arrays. Another interesting application for the algorithm would be
Josephson-junction systems coupled to quasi-particle dissipation
\cite{AES}.

\acknowledgements

We acknowledge support by the Swiss National Science Foundation and
NSF grant PHY99-07949, as well as helpful discussions with
S. Chakravarty, E. Demler and D. Fisher. Part of this work was completed at the Kavli Institute for
Theoretical Physics, UCSB. We are also grateful for the
hospitality of the Aspen Institute of Physics. The calculations have been performed on the Asgard and Hreidar Beowulf clusters at ETH Z\"urich, using the ALPS library \cite{ALPS}.

\appendix

\section{Critical exponents of the ICFP \label{appA}}

The behavior of the two-junction system in the ICFP region at intermediate energies is determined by the critical properties of the 
unstable fixed point. Here we derive the critical exponents and principal directions of the RG in the weak and strong coupling regimes. 

\subsection{Weak Coupling}

The weak coupling RG equations are given by
\begin{eqnarray}
\frac{dJ}{dl}&=&-J u+\frac{R}{R_Q} J J_+,\label{eq14aa}\\
\frac{dJ_+}{dl}&=&-J_+ w+\frac{r}{R_Q} J^2,
 \label{eq14ba}
\end{eqnarray}
where
\begin{equation}
u=\frac{R+r}{R_Q}-1, \hspace{5mm} w=\frac{2R}{R_Q}-1.
\label{eq14.5a}
\end{equation}
Near the unstable fixed point,
\be
J^*=\frac{R_Q}{\sqrt{rR}}\sqrt{u w},\hspace{5mm}  J^*_+=\frac{R_Q}{R}u,
\ee
we can linearize Eqs. (\ref{eq14aa}) and (\ref{eq14ba}) by
writing
\be
J=J^*+j, \hspace{5mm} J_+=J_+^*+j_+,
\label{eq16}
\ee
and thus obtain
\be
\frac{d}{dl}\l(\ba{c} 
j \\ j_+ \ea \r) = \l(\ba{cc} 0 & \sqrt{\frac{R}{r}}\sqrt{wu} \\ 
                          2\sqrt{wu}\sqrt{\frac{r}{R}}  &  -w \ea \r) \l(\ba{c} 
j \\ j_+ \ea \r) .
\label{eq17}
\ee
The eigenvalues and eigenvectors of Eq.~(\ref{eq17}) give the relevant
direction and exponent, and also the irrelevant direction and its
decay. For the relevant direction one finds
\begin{eqnarray}
\lambda_+&=&\frac{1}{2}\l(-w+\sqrt{w^2+8uw}\r) \\
(j,\,j_+)&=&\l(1,\,\frac{1}{2}\sqrt{\frac{r}{R}}\l(-\sqrt{\frac{w}{u}}+\sqrt{\frac{w}{u}+8}\r)\r),
\label{eq18}
\end{eqnarray}
and for the irrelevant direction
\begin{eqnarray}
\lambda_-&=&\frac{1}{2}\l(-w-\sqrt{w^2+8uw}\r) \label{eq19a} \\
(j,\,j_+)&=&\l(1,\,\frac{1}{2}\sqrt{\frac{r}{R}}\l(-\sqrt{\frac{w}{u}}-\sqrt{\frac{w}{u}+8}\r)\r).\hspace{5mm}
\label{eq19b}
\end{eqnarray}
Near the unstable fixed point we expect
\be
j\sim a T^{-\lambda_+}+b T^{-\lambda_-} , \hspace{5mm}  j_+\sim a_+ T^{-\lambda_+}+b_+ T^{-\lambda_-} ,
\label{eq21}
\ee
where $a$ and $b$ are determined by the initial  $J^{(0)}$
and $J_+^{(0)}$. At very low $T$ we are close to the $J=J_+=0$ fixed point and have
\be
J\sim T^{u}, \hspace{5mm} J_+\sim T^{w}.
\label{eq22}
\ee
If the system is superconducting, which means that initially $J^{(0)}$
and $J_+^{(0)}$ are in the region in which they grow, then
Eq.~(\ref{eq21}) still holds for intermediate
temperatures, but as soon as the J's are far from the ICFP, the system
crosses over to the strong coupling regime, where
\be
\zeta\sim T^{\frac{R_Q(R+r)}{2Rr+R^2}-1},
\label{eq24}
\ee
as follows from Eq.~(\ref{w20b}).

\subsection{Strong coupling regime}

The RG equations for the strong coupling case are
\begin{eqnarray}
\frac{d\zeta}{dl}&=&-\zeta\overline{u}+\frac{R R_Q}{R^2+2Rr} \zeta\zs,\\
\frac{d \zs}{dl}&=&-\zs\overline{v}+\frac{r R_Q}{R^2+2Rr} \zeta^2,
\label{ap1}
\end{eqnarray}
where
\begin{equation}
\overline{u}=\frac{R_Q (R+r)}{2Rr+R^2}-1, \hspace{5mm} \overline{w}=\frac{2R R_Q}{R^2+2Rr}-1.
\label{ap2}
\end{equation}
As in the weak coupling case we linearize these equations near the unstable fixed point
\be
\zeta^*=\frac{R^2+2rR}{R_Q \sqrt{rR}}\sqrt{\overline{u} \overline{w}},\hspace{5mm}
\zs^*=\frac{R^2+2rR}{R_Q R}\overline{u},
\label{ap3}
\ee
writing
\be
\zeta=\zeta^*+z, \hspace{5mm} \zs=\zs^*+z_D,
\ee
and thus obtain
\be
\frac{d}{dl}\l(\ba{c} 
z \\ z_D \ea \r) = \l(\ba{cc} 0 & \sqrt{\frac{R}{r}}\sqrt{\overline{w}\overline{u}} \\ 
                          2\sqrt{\overline{w}\overline{u}}\sqrt{\frac{r}{R}}  &  -\overline{w} \ea \r) \l(\ba{c} 
z \\ z_D \ea \r).
\label{ap4}
\ee
Eq.~(\ref{ap4}) is essentially identical to Eq.~(\ref{eq17}), for the
weak coupling case, except for the change of $u,\,w$ to
$\overline{w},\,\overline{u}$. Therefore all other results can be
copied from the previous section as well. The eigenvalues and
eigenvectors of the matrix in Eq.~(\ref{ap4}) are
\begin{eqnarray}
\lambda_+&=&\frac{1}{2}\l(-\overline{w}+\sqrt{\overline{w}^2+8\overline{u}\overline{w}}\r) \\
(z,\,z_D)&=&\l(1,\,\frac{1}{2}\sqrt{\frac{r}{R}}\l(-\sqrt{\frac{\overline{w}}{\overline{u}}}+\sqrt{\frac{\overline{w}}{\overline{u}}+8}\r)\r),\hspace{5mm}
\label{eq18s}
\end{eqnarray}
and for the irrelevant direction
\begin{eqnarray}
\lambda_-&=&\frac{1}{2}\l(-\overline{w}-\sqrt{\overline{w}^2+8\overline{u}\overline{w}}\r) \label{eq19sa}\\
(z,\,z_D)&=&\l(1,\,\frac{1}{2}\sqrt{\frac{r}{R}}\l(-\sqrt{\frac{\overline{w}}{\overline{u}}}-\sqrt{\frac{\overline{w}}{\overline{u}}+8}\r)\r).\hspace{5mm}
\label{eq19s}
\end{eqnarray}
Near the unstable fixed point we expect
\be
z\sim a T^{-\lambda_+}+b T^{-\lambda_-} , \hspace{5mm}  z_D\sim a_+ T^{-\lambda_+}+b_+ T^{-\lambda_-},
\ee
just as for $j$ and $j_+$ in Eq. (\ref{eq21}) for the weak coupling
case. 

\section{Approximate calculation of the NOR-FSC phase boundary \lb{appB}}

By using the result of App. \ref{appA} for the ICFP and the critical
exponents and irrelevant directions, Eqs.~
(\ref{eq19sa}) and (\ref{eq19s}), we can find an approximate implicit equation
which can be solved for the
critical $R$ as a function of $r$ and $\zeta_{0}$ (or $J_0$). The
calculation is identical for the two limits; we will demonstrate it
for the strong coupling limit.

The calculation is based on the simple assumption that for $r$, $R$ and $\zeta_0$ tuned to the critical values, the critical RG flow line between the point ($\xi=\xi_0$, $\xi_D$=0) and the fixed point ($\xi^*$, $\xi_D^*$) given in Eq. (\ref{i18}), is a straight line.
This is illustrated in Fig. \ref{appfig1}. The slope
of the line representing the critical manifold is 
\be
\tan\theta=|z_D/z|=\frac{1}{2}\sqrt{\frac{r}{R}}\l(\sqrt{\frac{\overline{w}}{\overline{u}}}+\sqrt{\frac{\overline{w}}{\overline{u}}+8}\r),
\ee
where $(z,z_D)$ is defined in Eq.~(\ref{eq19s}), and with $\overline{u}$
and $\overline{w}$ (which are functions of $r$ and $R$) given in
Eq. (\ref{ap2}). 
The implicit equation for $R_c$ is 
\be
\zs*=\l(\zeta_0-\zeta^*\r)\tan\theta .
\ee
More explicitly it becomes
\begin{align}
&\frac{R^2+2rR}{R_Q R}\overline{u} =\nonumber\\
&\l(\zeta_0-\frac{R^2+2rR}{R_Q
  \sqrt{rR}}\sqrt{\overline{u}
  \overline{w}}\r)\frac{1}{2}\sqrt{\frac{r}{R}}\l(-\sqrt{\frac{\overline{w}}{\overline{u}}}-\sqrt{\frac{\overline{w}}{\overline{u}}+8}\r).
\end{align}

\begin{figure}
\includegraphics[width=7.5cm]{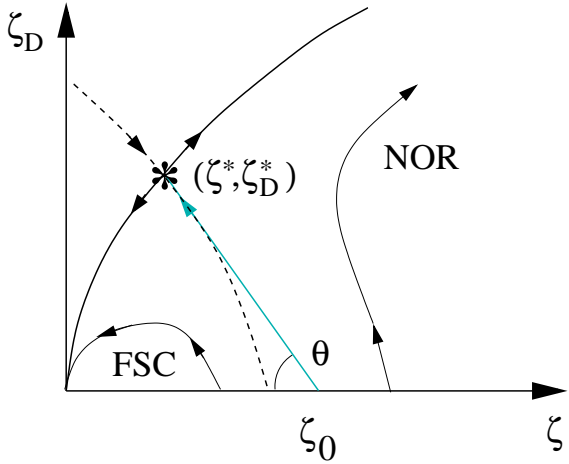}
\caption{We calculate the critical $R$ as a function of $r$ and
  $\zeta_0$ by assuming that the flow from the initial state,
  $(\zeta_0,\,0)$ to the fixed point, $(\zeta^*,\,\zs^*)$ is a
  straight line (dashed-dotted line). This turns out to be a reliable approximation
  throughout the ICFP region. \lb{appfig1}}
\end{figure}

\section{Solution of the circuit in Fig. \ref{ICFPcircuit} \lb{appC}}

The effective lead-to-grain resistance is obtained by solving the
circuit in Fig. \ref{ICFPcircuit} while replacing the Ohm-meter with a
current source providing a current $I$. The following equations
express zero potential drop along the two loops in the circuit:
\begin{align}
&\l(I+I_1\r)R+\l(I+I_1-I_2\r)r+I_1\alpha\zeta^2+\l(I_1-I_2\r)\alpha\zs^2=0,\\
&I_2 R+I_2\alpha\zeta^2-\l(I_1-I_2\r)\alpha\zs^2=\l(I+I_1-I_2\r).
\lb{C1}
\end{align}
The solution for $I_1$ and $I_2$ is
\begin{eqnarray}
I_1&=&-I\cdot\frac{r\l(2R+\alpha\zeta^2\r)+R\l(R+\alpha\zeta^2+\alpha\zs^2\r)}{\l(R+\alpha\zeta^2\r)\l(2r+R+\alpha\zeta^2+2\alpha\zs^2\r)},\lb{C2a}\hspace{3mm}\\
I_2&=&I\cdot\frac{r\alpha\zeta^2-R\alpha\zs^2}{\l(R+\alpha\zeta^2\r)\l(2r+R+\alpha\zeta^2+2\alpha\zs^2\r)},
\lb{C2b}
\end{eqnarray}
and the potential drop on the junction connecting the grain B and the lead
C is
\be
V_{BC}=-\l(I_1\alpha\zeta^2+\l(I_1-I_2\r)\alpha\zs^2\r).
\lb{C3}
\ee
Substituting Eqs.~(\ref{C2a}) and (\ref{C2b}) into Eq.~(\ref{C3}), and dividing by $I$, we obtain the
effective resistance measured in the Monte Carlo calculations:
\be
R_\text{eff}=\frac{\l(R^2+2rR\r)\l(\zeta^2 + \zs^2\r)
  +(R+r)\alpha\zeta^2\l(\zeta^2 + 2\zs^2\r)}{\l(R/\alpha + \zeta^2\r)\l(2r + R +\alpha\zeta^2 + 2\alpha\zs^2\r)}.
\lb{C4}
\ee


\begin{thebibliography}{99}

\bibitem{Refael1} G. Refael \textit{et al.}, Phys. Rev. B 68, 214515 (2003).
\bibitem{Chakravarty} S. Chakravarty, Phys. Rev. Lett. {\bf 49}, 681 (1982); J.~Fr\"ohlich and T.~Spencer, Commun. Math. Phys. {\bf 84}, 87 (1982). 
\bibitem{Bulgadaev} S. A. Bulgadaev, Sov. Phys. JETP Lett. {\bf 39}, 315 (1984).
\bibitem{Guinea} F. Guinea \textit{et al.}, Phys. Rev. Lett. {\bf 54}, 263 (1985
).  
\bibitem{Fisher&Zwerger} M. P. A. Fisher and W. Zwerger, Phys. Rev. B {\bf 32}, 6190 (1985).
\bibitem{Schmid} A. Schmid, Phys. Rev. Lett. {\bf 51}, 1506 (1983).
\bibitem{Chakravarty&Ingold1} S. Chakravarty \textit{et al.}, Phys. Rev. Lett. {
\bf 56}, 2303 (1986); S. Chakravarty \textit{et al.}, Phys. Rev. B {\bf 37}, 328
3 (1988). 
\bibitem{Panyukov&Zaikin} S. V. Panyukov and A. D. Zaikin, Phys. Lett. A {\bf 124}, 236 (1988); W. Zwerger, Physica B {\bf 152}, 236 (1988).
\bibitem{Korshunov1} S. E. Korshunov, Zh. Eksp. Teor. Fiz. {\bf 95}, 1058 (1989); S.~E.~Korshunov, Europhys. Lett. {\bf 9}, 107 (1989).
\bibitem{Haviland} E. Chow, P. Delsing and D. B. Haviland, Phys. Rev. Lett. {\bf 81}, 204 (1998).
\bibitem{Kobayashi} Y. Takahide , A. Kanda, Y. Ootuka, S.-I. Kobayashi, Phys. Rev. Lett. {\bf 85}, 1974 (2000).
  \bibitem{Refael2} G. Refael \textit{et al.}, to be published.
\bibitem{Tewari} S. Tewari \textit{et al.}, cond-mat/0407308; cond-mat/0501219.
\bibitem{junction} P. Werner and M. Troyer, Phys. Rev. Lett. {\bf 95}, 060201 (2005). 
\bibitem{Caldeira} A. O. Caldeira and A. J. Leggett, Ann. Phys. {\bf 149}, 374 (1983).
\bibitem{Schoen} G. Sch\"on and A. D. Zaikin, Phys. Rep. {\bf 198}, 237 (1990). 
\bibitem{Kivelson-Chakravarty} S. Chakravarty, G. Ingold, S. Kivelson, and G. Zimanyi,
Phys. Rev. B {\bf 37}, 3283 (1988).
\bibitem{AES} V. Ambegaokar, U. Eckern and G. Sch\"on, Phys. Rev. Lett. {\bf 48}, 1745 (1982).
\bibitem{ALPS} M. Troyer {\it et al.}, Lecture Notes in Computer Science {\bf 1505}, 191 (1998); F. Alet \textit{et al.}, J. Phys. Soc. Jpn. Suppl. {\bf 74}, 30 (2005); \url{http://alps.comp-phys.org/}.



\end{thebibliography}
\end{document}